\documentclass[aps,pre,twocolumn,superscriptaddress,amsmath,amssymb,a4paper,longbibliography]{revtex4-1}
\usepackage[utf8]{inputenc}

\usepackage{amsmath}
\usepackage{amssymb,amsfonts,latexsym}
\usepackage{bm}
\usepackage[mathcal]{euscript}
\usepackage{graphicx}
\usepackage{epsfig}
\usepackage{xfrac}

\usepackage{color}
\usepackage{pifont,wasysym,marvosym}
\usepackage{textcomp}
\usepackage{comment}
\usepackage{epstopdf}
\usepackage{hyperref}

\begin{document}
\title{A characteristic lengthscale causes anomalous size effects and boundary programmability in mechanical metamaterials}

\author{Corentin Coulais}
\affiliation{AMOLF, Science Park 104, 1098 XG Amsterdam, The Netherlands}
\affiliation{Huygens-Kamerlingh Onnes Lab, Universiteit Leiden, PObox 9504, 2300 RA Leiden, The Netherlands}
\affiliation{Van der Waals – Zeeman Institute, Institute of Physics, Universiteit van
Amsterdam, Science Park 904, 1098 XH Amsterdam, The Netherlands}
\author{Chris Kettenis}
\affiliation{Huygens-Kamerlingh Onnes Lab, Universiteit Leiden, PObox 9504, 2300 RA Leiden, The Netherlands}
\author{Martin van Hecke}
\affiliation{AMOLF, Science Park 104, 1098 XG Amsterdam, The Netherlands}
\affiliation{Huygens-Kamerlingh Onnes Lab, Universiteit Leiden, PObox 9504, 2300 RA Leiden, The Netherlands}

\maketitle

\textbf{The architecture of mechanical metamaterials is designed to harness  geometry \cite{Lakes_poisson,Grima_squares,Kadic_pentamode,Maha_miura,Wegener_Milton,Wegener_cloak},
nonlinearity \cite{Mullin_holey,Motter_negcompress,Shim_buckliball,Coulais_Metabeam,Coulais_nonreci} and topology \cite{KaneLubensky,Vitelli_solitons,Huber_reviewtopological,Meeussen_gears,Coulais_nonreci} to obtain advanced functionalities such as shape morphing~\cite{Mullin_holey,Shim_buckliball,Waitukaitis_origami,Silverberg_squaretwist,Coulais_metacube,Maha_curvature,Overvelde_origami1,Overvelde_origami2}, programmability~\cite{Silverberg_miura,Florijn_biholar,Coulais_metacube} and one-way propagation~\cite{Vitelli_solitons,Huber_reviewtopological,Coulais_nonreci}.
While a purely geometric framework successfully
captures the physics of small systems under idealized conditions, large systems or heterogeneous driving conditions remain essentially unexplored.
Here we uncover strong anomalies in the mechanics of a broad class of metamaterials, such as auxetics~\cite{Grima_squares,Bertoldi_poisson,Wegener_Milton}, shape-changers~\cite{Waitukaitis_origami,Silverberg_squaretwist,Coulais_metacube,Maha_curvature,Overvelde_origami1,Overvelde_origami2} or topological insulators~\cite{KaneLubensky,Vitelli_solitons,Coulais_nonreci,Meeussen_gears}:
a non-monotonic variation of their stiffness with system size, and
the ability of textured boundaries to completely alter their properties.
These striking features stem from the competition between rotation-based deformations---relevant for small systems---and
ordinary elasticity, and are controlled by a characteristic length scale
which is entirely tunable by the architectural details.
Our study provides new vistas for designing,  controlling and programming the
mechanics of metamaterials in the thermodynamic limit.}

A central strategy for the design of metamaterials leverages the notion of a mechanism, which is a collection of rigid elements linked by completely flexible hinges, designed to allow for a collective, free rotational motion of the elements. Mechanism-based metamaterials borrow the geometric design of mechanisms, but instead of hinges feature flexible parts which connect stiffer elements~\cite{Milton_tensors,Grima_squares,Wegener_Milton,Florijn_biholar,Waitukaitis_origami,Coulais_metacube,Coulais_nonreci,Kadic_pentamode,KaneLubensky,Vitelli_solitons,Meeussen_gears,Shim_buckliball,Waitukaitis_origami,Coulais_metacube,Silverberg_miura,Silverberg_squaretwist,Lechenault_OriLengthscale}. The tacit assumption is then
that the low-energy deformations of such metamaterials are similar to the free motion of the underlying mechanism, and
the ability to control deformations by geometric design
is the foundation for the unusual mechanics of a wide variety of mechanical  metamaterials. Such mechanism-based metamaterials have mostly been studied for small systems and for homogeneous loads,
where the response indeed closely follows that of the underlying mechanism.
However, the physics of large systems, or for inhomogeneous boundary conditions, remains largely unexplored.

We first illustrate that deformations of mechanism-based metamaterials deviate from those of their underlying mechanism  under inhomogeneous forcing. Specifically, we
consider point forcing of a paradigmatic  metamaterial (Fig.~1a), which is based on a  mechanism consisting of counter-rotating hinged squares (Fig.~1b) \cite{Grima_squares,Mullin_holey,Bertoldi_poisson,Shim_buckliball,Florijn_biholar,Coulais_Metabeam}. Whereas the local deformations mimic that of the underlying mechanism,
at larger scales, we observe that the counter-rotations slowly decay away from the boundary (Fig.~1c).  {This indicates elastic distortions of the underlying rotating square mechanism, where no such decay can occur.}
In this example, 2D effects complicate the physics, and we therefore focus on quasi-1D meta-chains, consisting of $2\times N$ square elements of diagonal $L$ linked at their tips (Fig.~2a-b); whenever convenient, we will express lengths in units of $L$. 
{We measure the linear response of these samples by forcing the 
 outer horizontal joints.}
Surprisingly, both
experiments and finite element (FEM) simulations show an exponential decay of the mechanism-like rotations away from the boundary  when the meta-chain is stretched or compressed (Fig.~2c).
{This spatial decay defines} a novel characteristic length $n^*$ (Fig.~2c-inset), and suggests that elastic distortions of the underlying mechanism are a general feature of mechanism-based metamaterials.

 \begin{figure}[t!]
\includegraphics[width=.99\columnwidth,trim=0cm 0cm 0cm 0cm]{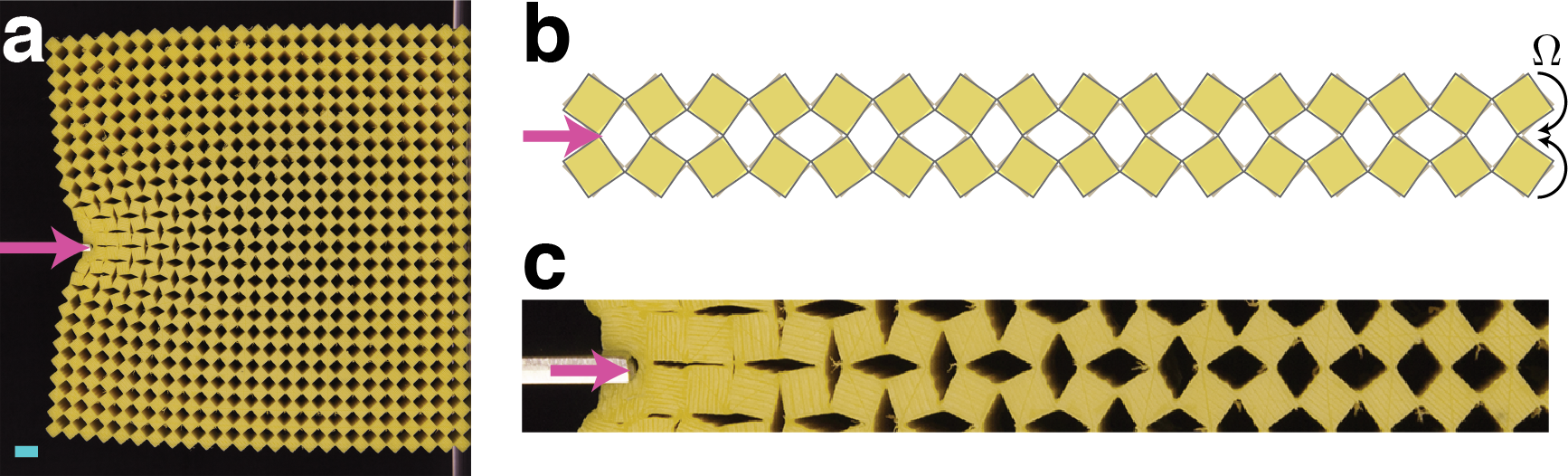}

\vspace{-0.3cm}
\caption{{\bf Mechanism-based metamaterials.} (a) A paradigmatic example of a mechanism based metamaterial consisting
of rubber slab patterned with a regular array of holes \cite{Grima_squares,Mullin_holey,Bertoldi_poisson,Shim_buckliball,Florijn_biholar,Coulais_Metabeam}.
Point-indentation excites a characteristic diamond-platter pattern near the tip and more smooth deformations further away (scale bar is $9$ mm).
(b) The rotating squares mechanism \cite{Grima_squares} consists of counter rotating, hinged rigid squares and underlies the design of the metamaterial shown in panel (a). The deformation from the symmetric state can be specified by a single angle $\Omega$.
(c) A zoom-in reveals that the deformation field of the mechanism-based metamaterial is highly textured, with the rotation $\Omega$ slowly decaying away from the boundary.}
\label{fig1}
\end{figure}

A first striking consequence of these distortions emerges when probing
the effective stiffness of mechanism-based metamaterials as function of system size.
While for elastic continua the effective spring constant or stiffness is inversely proportional to the linear size~\cite{Landau_Lifschitz},
experiments and finite element (FEM) simulations of meta-chains reveal remarkable deviations from this behaviour.
For small systems we find that the stiffness $k_o$ for odd $N$ is much larger than the stiffness $k_e$ for even $N$. Moreover, while $k_o$ decays monotonously with $N$, $k_e$ initially increases with $N$.
Eventually the stiffness $k_e$ peaks at length $n_p$, and for larger $N$, $k_e$ approaches $k_o$ and both decay with system size  (Fig.~2d). This anomalous size dependence is a robust feature---{we have numerically determined the size dependent stiffness for the 2D metamaterial shown in Fig.~1, as well as a 3D generalization of these \cite{Coulais_metacube}, and find that these exhibit a similar peak in stiffness (see Extended Data Figure~\ref{ED:2D3D}.)}


 \begin{figure}[t!]
\includegraphics[width=.99\columnwidth,trim=0cm 0cm 0cm 0cm]{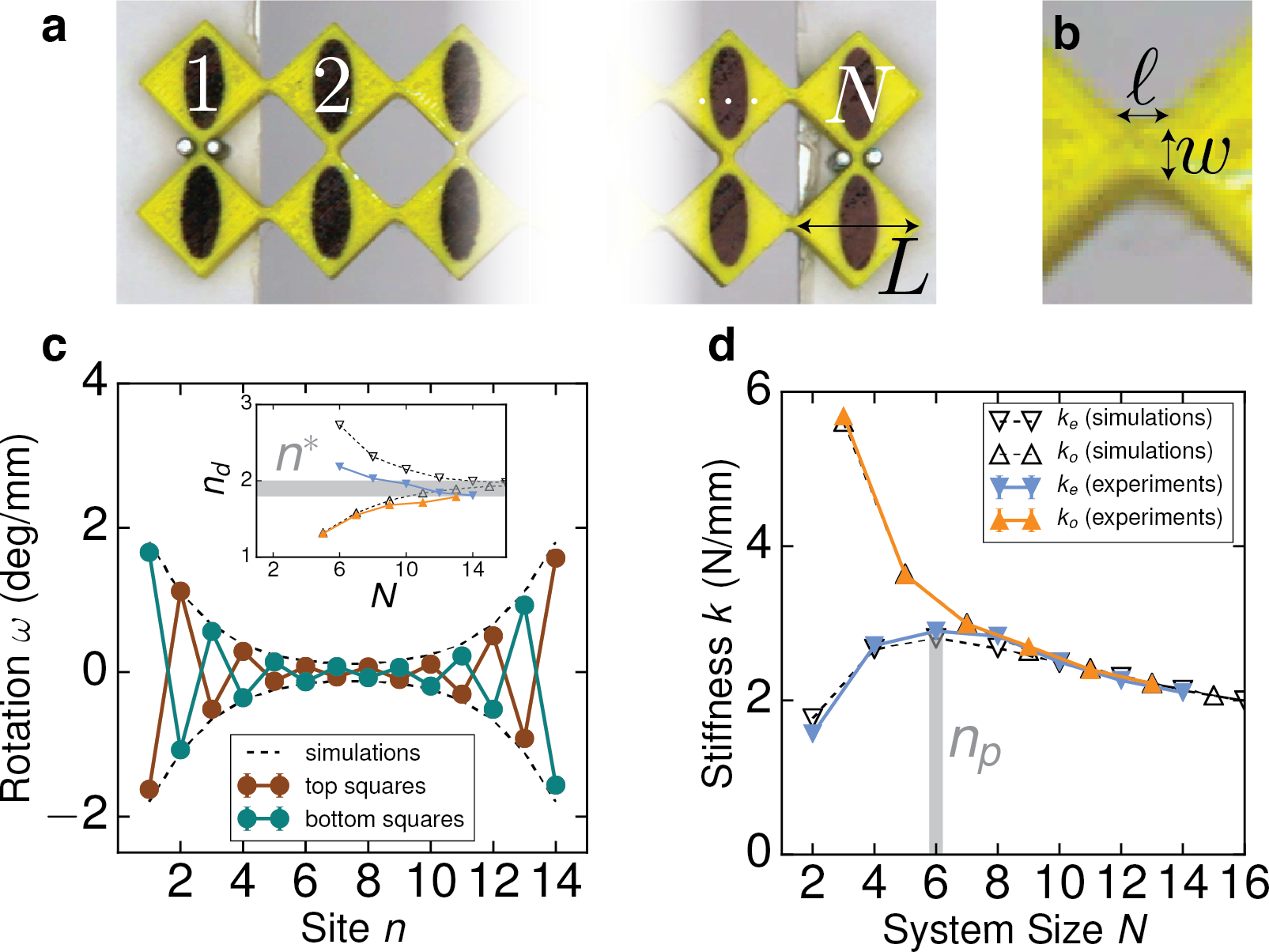}

 \vspace{-0.3cm}
\caption{{\bf Anomalies in the stiffness and deformations of meta-chains.} (a) 3D printed meta-chain of length $N$, thickness $H=7.5$ mm and square {diagonal} $L=17$ mm; the black ellipses are used for tracking positions and rotations. (b) Hinge geometry defined by $\ell$ and $w$. (c) Rotation  field for a meta-chain of length $N=14$ (See Methods). Maroon (blue) symbols denote the data obtained from the upper (lower) squares. Inset:
The decay length converges to a well defined value $n^*=1.9\pm 0.1$ in the large system size limit. (d) Stiffness as function of $N$ {(See Methods)}. Coloured symbols denote experiments for odd ($k_o$; orange) and even ($k_e$; blue) meta-chains, and dashed curves denote FEM simulations, for  $\ell=1.7$ mm, $w=1.7$ mm.
The stiffness $k_e$ peaks at  $n_p=6.0\pm 0.1$.
}
\label{fig2}
\end{figure}

The stiffness anomaly reflects the hybrid nature of mechanism-based metamaterials,
as can be seen by comparing two simple models.
While a chain consisting of $N$ unit springs of stiffness $\kappa$ in series has
a global spring constant $k$ that is inversely proportional to the system size $N$: $k=\kappa/N$, the stiffness of a rotating squares chain where all hinges are dressed by torsional springs of stiffness $C_b$~\cite{Florijn_biholar,Waitukaitis_origami,Coulais_nonreci} does not decay with $N$ (See Supplementary Information). Specifically, for even $N$, the local rotation $\Omega$ and globally applied deformation $u$ are of the same order, and the spring constant $k_e \sim N$ --- longer chains are thus stiffer in this model. For odd $N$, the counter rotating motions cancel in leading order, so that $\Omega \ggg u$ and $k_o$ diverges (see Supplementary Information).
Hence,  whereas the total deformation in a spring chain is evenly distributed over all elastic elements, such homogeneity breaks down for mechanisms, precisely because of the counter-rotations.
The response of flexible, mechanism-based metamaterials {hybridises}
pure mechanism-like and homogeneous elastic deformations, leading to a crossover from a mechanism-dominated, inhomogeneous regime for small systems to a homogeneous
elastic regime for larger system sizes.

Both $n^*$ and $n_p$ reveal this crossover, but we note that their values differ.
To understand what sets these values and untangle
their relation, we
%
%
consider a hybrid dressed mechanism where the hinges
are subject to bending, stretching and shear, with stiffnesses $C_b$, $k_j$ and $C_s$ respectively.
Stretching and shear introduce deformations that compete with the purely counter rotating mode of the underlying  mechanism.
The equations that govern mechanical equilibrium
are controlled by the dimensionless ratios (see Supplementary Information):
\begin{equation} \label{dimpar}
	\alpha = \left( 1 {+} \frac{L}{\ell} \right)^2 \frac{C_s}{4C_b}; \quad \beta = \frac{k_j L^2}{4C_b}~,
\end{equation}
which tune the relative elastic penalties of mechanism-preserving and mechanism-distorting deformations. The purely torsional model corresponds to the limit where both the stretching and shear stiffnesses are much larger than the bending stiffness, that is ($\alpha,\beta \rightarrow \infty$).
We have checked that solutions to this model for appropriate values of $\alpha$ and $\beta$ show excellent agreement with the experimental results (see Methods and Extended Data Figures~\ref{ED:hinges},\ref{ED:model}): dressing the mechanism with elastic hinge interactions
is an effective approach to describe mechanism-based metamaterials.

\begin{figure}[t!]
\includegraphics[width=.99\columnwidth,trim=0cm 0cm 0cm 0cm,clip]{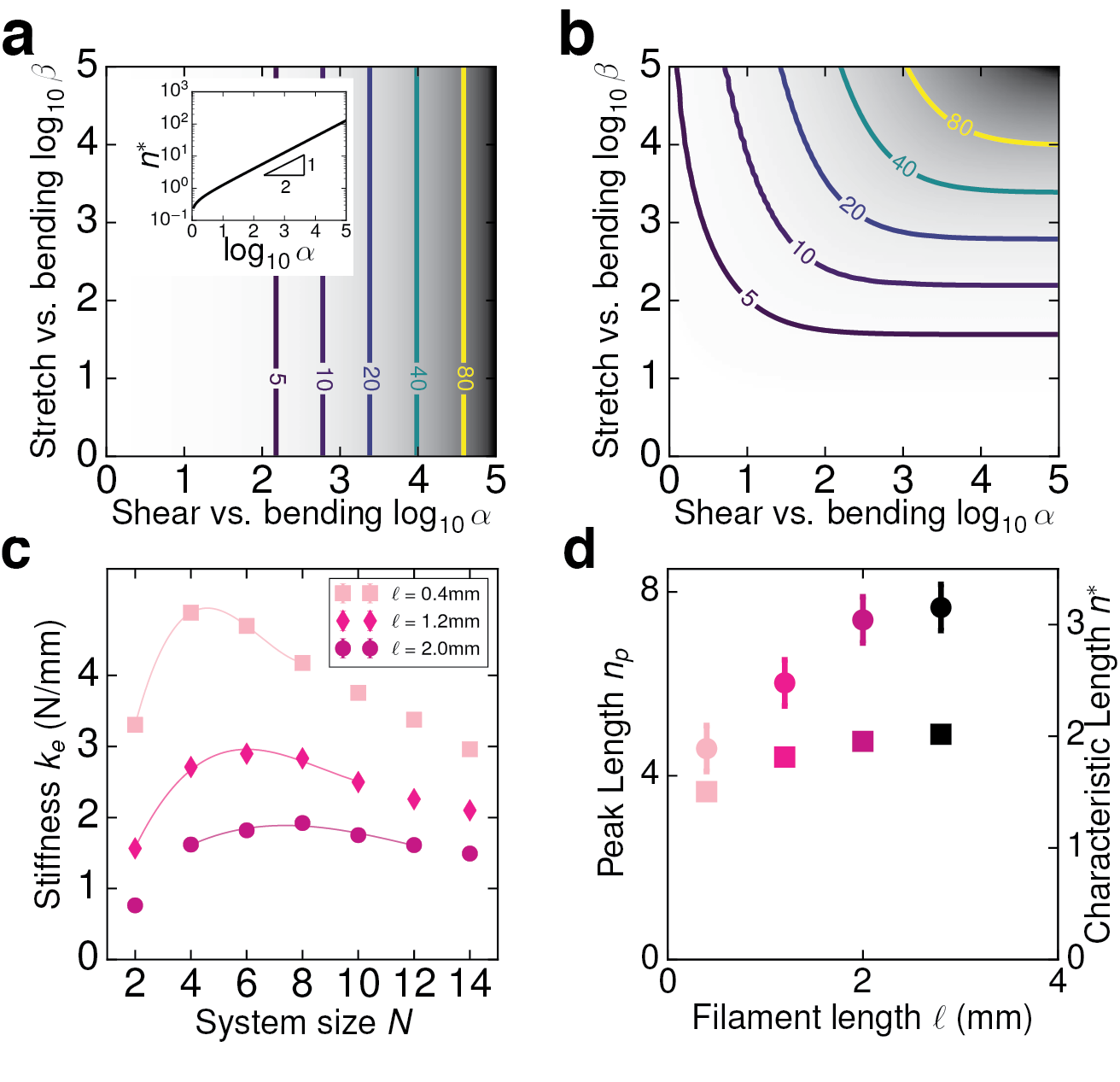}

 \vspace{-0.7cm}
\caption{{\bf Characteristic scales:} (a-b) Contour plots of $n^*$ (a) and $n_p$ (b) vs. the shear-to-bending ratio $\alpha$ and  the stretch to bending ratio $\beta$ computed for the hybrid mechanism (see methods). (a-inset) The characteristic length
$n^*$ scales as the square root of $\alpha$. (c) Stiffness vs. system size in experiments with different filament length $\ell$. We fit a cubic function (continuous curves) to the data near the peak of $k_e$, allowing us to estimate $n_p$ to within $\pm 0.1$. (d). Corresponding location of the lengthscales  $n_p$ (disks) and $n^*$ (squares) vs. filament length $\ell$.}
\end{figure}

The competition between mechanism-preserving and mechanism-distorting deformations
controls the characteristic length scale. To show this, we  vary the control parameters $\alpha$ and $\beta$, and determine  $n^*$ and $n_p$. When mechanism-like deformations are energetically cheap (large $\alpha$, $\beta$), both $n^*$ and $n_p$ diverge, whereas when rotations are energetically expensive (small $\alpha$, $\beta$),
the lengths $n^*$ and $n_p$ become small  (Fig.~{3}a-b).
Experimentally, we can leverage this connection to vary and control the length scale, as
the relative costs of the mechanism-preserving and mechanism-distorting deformations are controlled
by the hinge geometry. To demonstrate this, we have varied the experimental hinge length {$\ell$} to push the stiffness ratio's $\alpha$ and $\beta$ up, and we
find that increasing $\ell$  indeed leads to an increase of both $n^*$ and $n_p$  (Fig.~{3}c-d).

Strikingly, $n^*$ is independent of $\beta$ whereas $n_p$ depends on both $\alpha$ and $\beta$, and as we will show below, also on the boundary conditions. The variation of
$n^*$ with $\alpha$ can be understood  from
the competition between the energy cost $\sim N C_b u^2$ of purely counter rotating deformations, and the energy cost $\sim C_s/N u^2$ of a shear-induced gradient of these rotations.
Balancing these terms yields a characteristic {length}
$n^*~\sim \sqrt{C_s/C_b} ~\sim \sqrt{\alpha}$, consistent with our data (Fig.~{3}a inset).
We note that exactly solving the underlying equations of the dressed mechanism make this argument rigorous (see Supplementary Information).

In contrast, the length scale $n_p$ depends on both $\beta$ and the boundary conditions. To probe this boundary dependence, we consider
boundary conditions where we  independently control the forces $F$ (red) and $F'$ (blue) at alternating locations at the edge of the chain, by setting $F'=\tfrac{\lambda}{2} F$ (Fig.~4a) ; so far, we considered $\lambda=0$. The intrinsic lengthscale $n^*$ is insensitive to the choice of boundary conditions, but the boundary hybridisation factor $\lambda$ allows to control $n_p$ over a wide range (Fig.~4b), by tuning the magnitude of the rotational field (Fig.~4c). To illustrate that this sensitivity to boundary conditions is relevant for a wide class of mechanism-based metamaterials,
we  consider a topological metamaterial which exhibits one way motion amplification ~\cite{Coulais_nonreci} (Fig.~4d).
For a hybrid mechanism where the hinges are dressed with torsional and stretch interactions, the boundary conditions control the hybridisation of mechanism-like and ordinary elastic deformations. Surprisingly, whereas {in the mechanism-limit} deformations are located near the right boundary, so that forces/displacements excited from the left are amplified, manipulation of the boundary conditions allows to tune the gain of the displacement amplification over a giant---$80$dB---range (Fig.~4e) and to excite deformations that can be localized near the left edge, near the right edge, or near both boundaries (Fig.~4f).
Hence, the introduction of  finite energy distortions alleviates topological protection and allows boundary programmability.

\begin{figure}[t!]
 \includegraphics[width=.99\columnwidth,trim=0cm 0cm 0cm 0cm,clip]{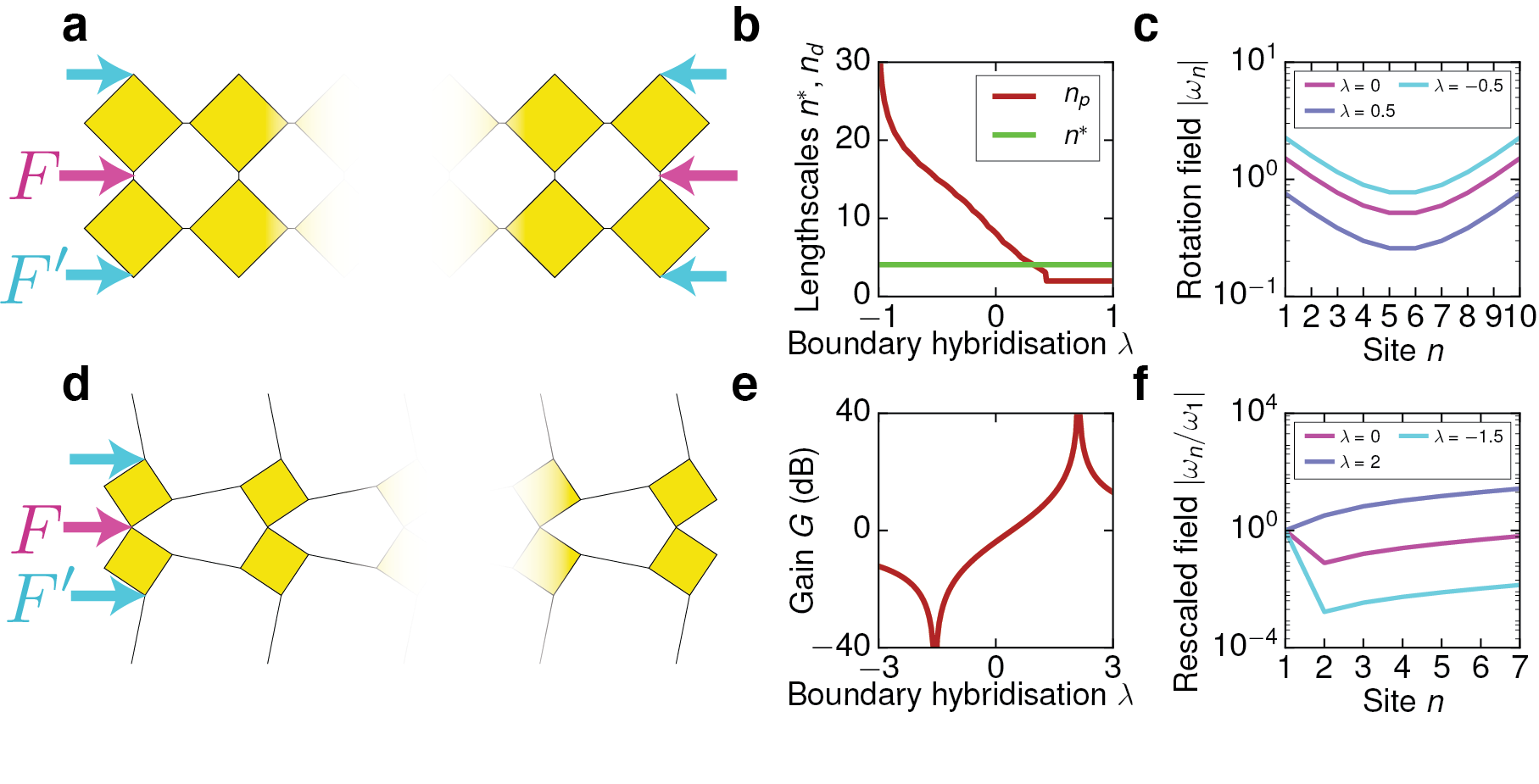}
 
 \vspace{-0.7cm}
\caption{{\bf Sensitivity to boundary conditions.} (a) Meta-chain with
imposed forces $F$ (violet) and $F':= \tfrac{\lambda}{2} F$ (blue). (b) Length scales vs $\lambda$, illustrating that $n^*$ is an intrinsic feature while $n_p$ can be tuned by the boundary conditions. (c) The amount of rotations depends strongly on the boundary conditions. (d) Topological chain (See Supplementary Information for the theoretical description). (e) The displacement amplification gain  $G$ strongly depends on the boundary condition hybridisation factor $\lambda$. The gain is defined as $G = 20 \log_{10} \omega_N /\omega_1$, where $\omega_1$ ($\omega_N$) is the rotation of the most left (right) squares.
(f) Rotational field $\omega_n$ as function $\lambda$.}
\end{figure}

A physically appealing picture appears: mechanism-based metamaterials have an intrinsic length-scale $n^*$ that {depends on the geometric design and} diverges in the purely mechanism limit.
{Such length-scale quantifies the spatial extension of a} soft mode, which localizes near inhomogeneities such as boundaries. Whether this mode is excited depends on the boundary conditions. For the case of the meta-chain, if we choose boundary conditions which are compatible with the counter-rotating
texture of the underlying mechanism, i.e., $\lambda = -1$, the crossover length $n_p$ between mechanism-like and elastic behaviour diverges, whereas strongly incompatible boundary conditions lead to a rapid crossover to ordinary elastic behaviour.

We expect that most mechanism-based metamaterials, {including}
cellular metamaterials~\cite{Mullin_holey,Shim_buckliball,Florijn_biholar,Bertoldi_poisson,Coulais_Metabeam,Coulais_metacube,Coulais_nonreci}, {allosteric networks~\cite{Nagel_allostery,Wyart_allostery}}, gear-based metamaterials~\cite{Meeussen_gears} {and} origami~\cite{Silverberg_miura,Maha_miura,Waitukaitis_origami,Maha_curvature,Overvelde_origami1,Overvelde_origami2}, feature similarly large characteristic scales. Continuum descriptions need to encompass such large scales---in contrast to Cosserat-type descriptions of  random cellular solids governed by the bare cell size---as
well as the compatibility between the textures of the mechanism and the boundary \cite{Coulais_metacube}. We stress that proper hinge design is critical for maintaining functionality in large metamaterial samples, and we
suggest to explore hierarchical designs, with multiple small sub-blocks
connected via ''meta connectors'' that promote the propagation of the required mechanism in each block, thus ensuring that the functionality  survives elastic hybridization in the thermodynamic limit.

{\em Acknowledgements}. We thank J. Mesman for outstanding technical support and A. Al\`u, A. Meeussen and A. Souslov for insightful discussions. We acknowledge funding from the Netherlands Organization for Scientific Research through grants VICI No. NWO-680-47-609 (M.v.H.) and VENI NWO-680-47-445 (C.C).


%

\newpage
\section{Methods}
\subsection{Experiments}
We fabricate our samples by 3D printing a flexible polyethylene/polyurethane thermoplastic mixture (Filaflex by Recreus, Young's modulus $E=12.75$ MPa, Poisson's ratio $\nu\sim 0.5$). The samples are $7.5$ mm thick, initially made of $N=14$ rows of squares of {diagonal} $L=17$ mm, which are connected by ligaments of length $\ell=1.7$ mm and width $w=1.7$ mm (Fig.~2ab of the main text). We measure the stiffness of these samples by pinching the outer horizontal joints between two rods, which are positioned such that they tightly grip the joints---this boundary condition ensures that the rotational mode is strongly excited. The rods are attached to an uniaxial testing device equipped with a $100$ N load cell, which measures forces $F$ and displacements $\delta$ with $1$ mN and $10 ~\mu$m accuracy respectively, and with which we apply an external displacement from $\delta=-0.50$ mm (in compression) to $\delta=1.50$ mm (in extension). We focus on the linear response regime, and measure the stiffness $k$ in the displacement range $\delta\in[-0.25$ mm$, 1.25$ mm$]$ by using the linear coefficient of a $5^\textrm{th}$ order polynomial fit to the force-displacement $F$ vs. $\delta$ curve---details of the procedure are not crucial, as the data is very close to linear (See Extended Data Figure~\ref{ED:exp}a). To measure the variation of $k$ with $N$, we print $N=14$ samples, perform experiments, remove a pair of squares to obtain $N=13$, perform more experiment and so on.

We have marked these elements and record images with a high resolution CMOS camera (Basler acA2040-25gm; resolution $4$Mpx), which is triggered by the mechanical testing device. This allows us to measure rotations $\theta(n)$ with $1\times 10^{-1}$ deg accuracy vs. the displacement $\delta$. In the linear regime, $\theta(n)$ is proportional to $\delta$, and we determine the rotational rate $\omega(n)$ from a linear fit of the $\theta(n)$ vs. $\delta$ curves (See Extended Data Figure~\ref{ED:exp}b).


\subsection{Numerical Simulations}
For our static finite elements simulations, we use the commercial software \textsc{Abaqus/Standard} and we use a neohookean energy density as a material model, using a shear modulus, $G = 4.25$MPa and bulk modulus, $K=212$GPa (or equivalently a Young's modulus $E=12.75$~MPa and Poisson's ratio $\nu=0.49999$)
in plane strain conditions with hybrid quadratic triangular elements (abaqus type CPE6H). We perform a mesh refinement study in order to ensure that the thinnest parts of the samples where most of the stress and strain localized are meshed with at least four elements. As a result, the metamaterial approximately has from $3\times10^3$ to $6\times 10^4$ triangular elements, depending of the value of $N$.

\subsubsection{Simulation of the full metamaterial}
We simulate the full metamaterial by applying boundary conditions by pinching the most outer vertical connections as in the experiments. We impose a small displacement of magnitude $\delta=3\times 10^{-4}L$ to the structure and measuring the reaction force $F$. Given that such small displacement ensures the structure is probed in its linear response, we estimate the stiffness as $k=F/\delta$.

\subsubsection{Measurement of the hinge stiffnesses}
We measure the individual bending, stretch and shear stiffness by simulating two squares connected by one elastic ligament and applying three sorts of boundary conditions depicted in Extended Data Figure~\ref{ED:hinges}. To apply bending, stretching and shear boundary conditions, we define constraints for every node on the vertical diagonal of each square and assign their displacements to the motion of a virtual node, which is then displacement by a small amount $\delta=3\times 10^{-4}L$. We then extract the reaction forces $F_b$, $F_j$ and $F_s$, respectively on this virtual node to calculate the stiffnesses as follows
\begin{eqnarray}
C_b&=&\frac{L^2}{4}\frac{F_b}{\delta},\\
k_j&=&\frac{F_j}{\delta},\\
C_s&=&\frac{\ell^2}{2}\frac{F_s}{\delta}.\\
\end{eqnarray}


\clearpage
\section{Extended Data}
\setcounter{figure}{0}
\renewcommand{\figurename}{Extended Data Figure}

\begin{figure}[h!]
 \includegraphics[width=.9\columnwidth,trim=0cm 0cm 0cm 0cm,clip]{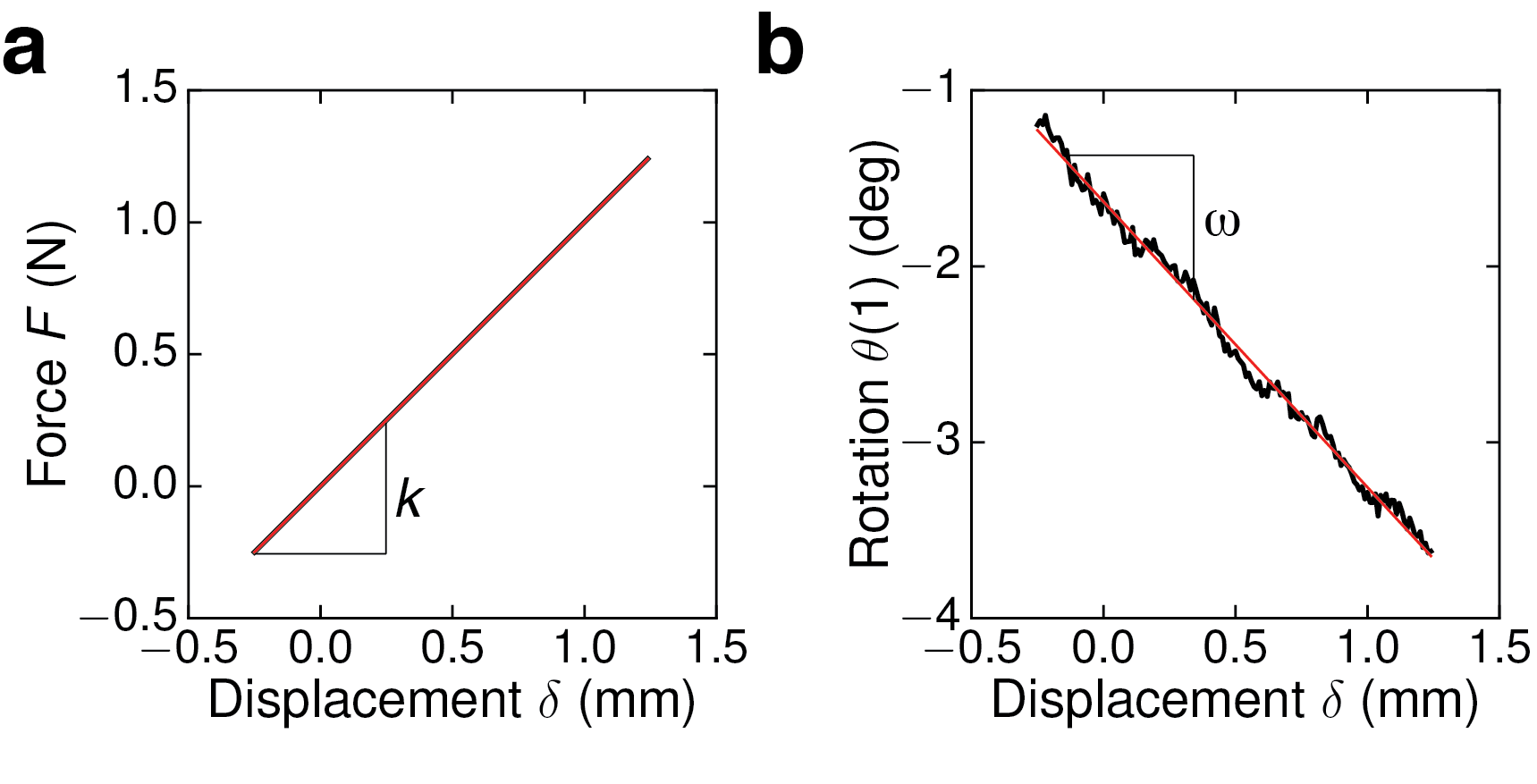}
\caption{{\bf Experimental determination of the stiffness and of the rotational rate.} (a) Force $F$ vs. displacement $\delta$ for a meta-chain of size $N=14$ (black curve). The stiffness $k$ is measured from the coefficients of a $5^\textrm{th}$ order polynomial fit (red line) to the data. (b) Rotation $\theta(1)$ of the bottom $n=1$ square vs. displacement $\delta$ (black curve). The rotational rate $\omega(1)$ is determined from a linear fit to the data (red line).}
\label{ED:exp}
\end{figure}

\begin{figure}[h!]
\includegraphics[width=.9\columnwidth,trim=0cm 0cm 0cm 0cm,clip]{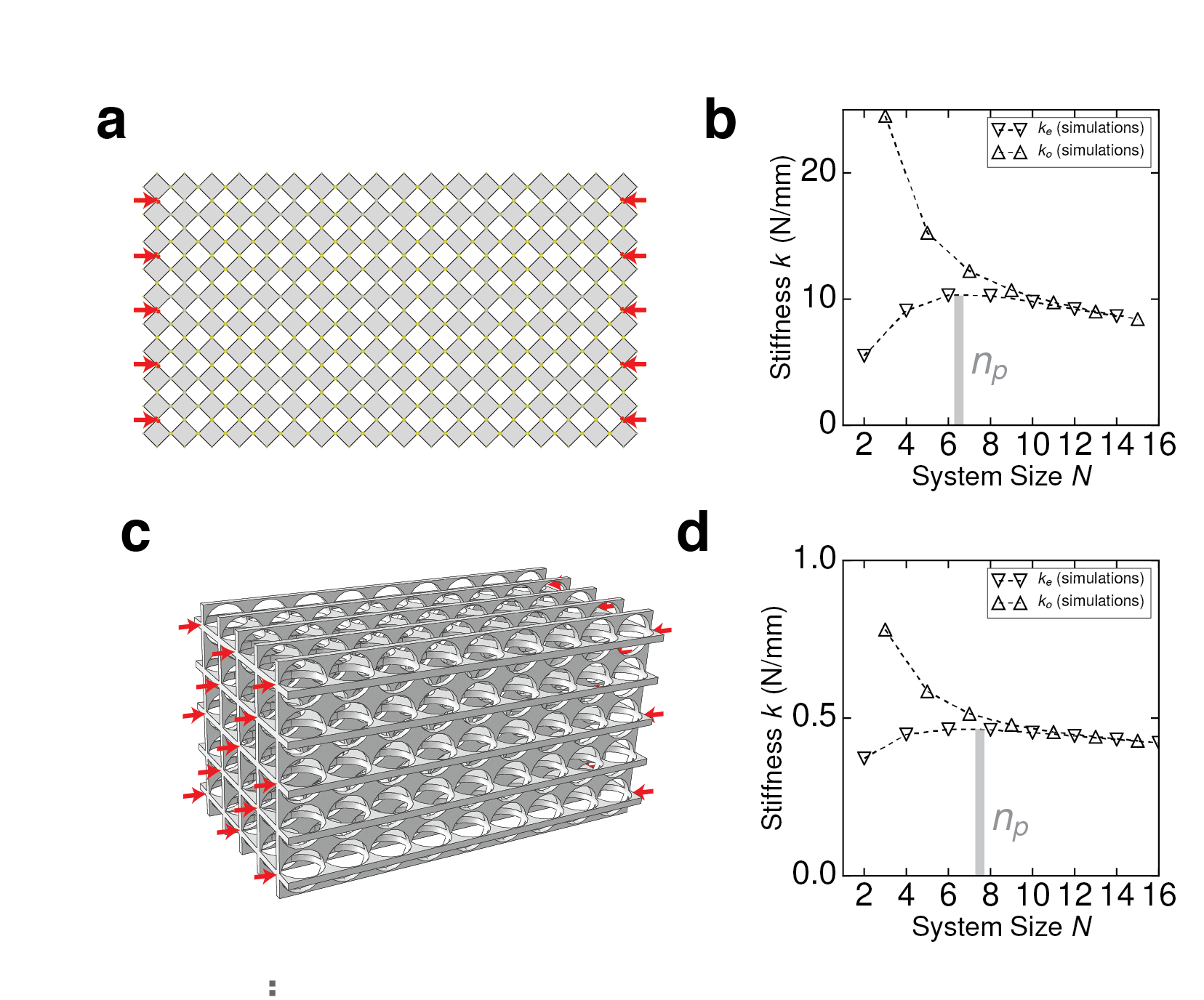}
\caption{{\bf Anomalous stiffness size dependence in 2D and 3D mechanical metamaterials.}
(a) 2D Metamaterial based on the 2D rotating square mechanism (see Fig. 1) under textured boundary conditions. (b) Stiffnesses $k_e$ and $k_o$ vs. system size $N$
obtained by numerical simulations. The geometric design and method of simulations are the same as for the meta-chain shown in Fig~.2. (c) 3D Metamaterial \cite{Coulais_metacube} under ``checkerboard'' textured boundary conditions. (d) Stiffnesses $k_e$ and $k_o$  vs. system size $N$ obtained by numerical simulations. The geometric design and simulation method follow those described in \cite{Coulais_metacube}, with the exception of the value of the struts width, $w$, which is chosen here twice as small.}
\label{ED:2D3D}
\end{figure}

\begin{figure}[h!]
	\centering
\includegraphics[width=0.9\columnwidth,trim=0cm 7cm 0cm 7cm]{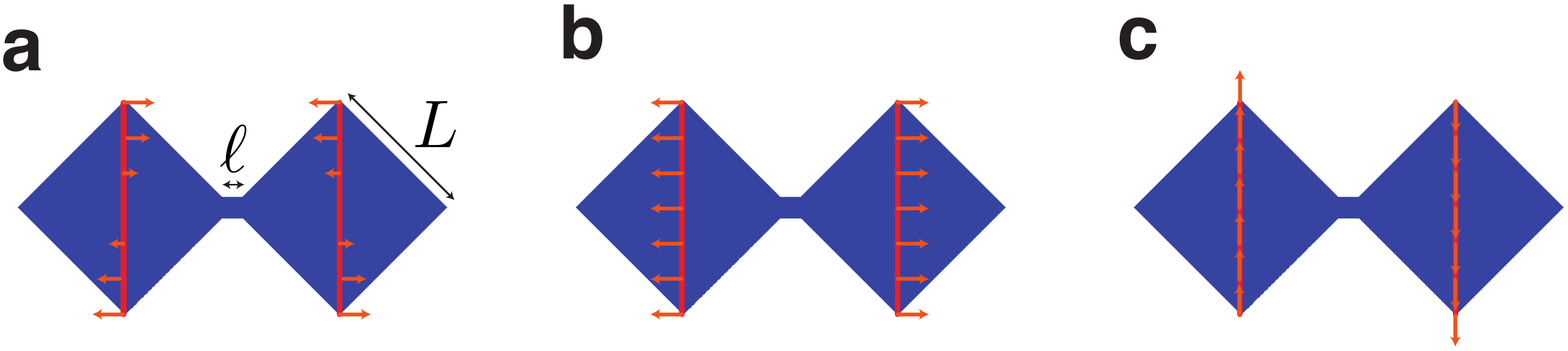}		
	\caption{{\bf FEM Simulation protocol to characterize the hinges}.  (a) Bending torsional stiffness $C_b$. (b) Stretching stiffness $k_j$. (c) Shear torsional stiffness $C_s$. The nominal applied relative displacement has a magnitude $3\times10^{-4} ~L$.}
	\label{ED:hinges}
\end{figure}

\begin{figure}[h!]
\includegraphics[width=.9\columnwidth,trim=0cm 0cm 0cm 0cm,clip]{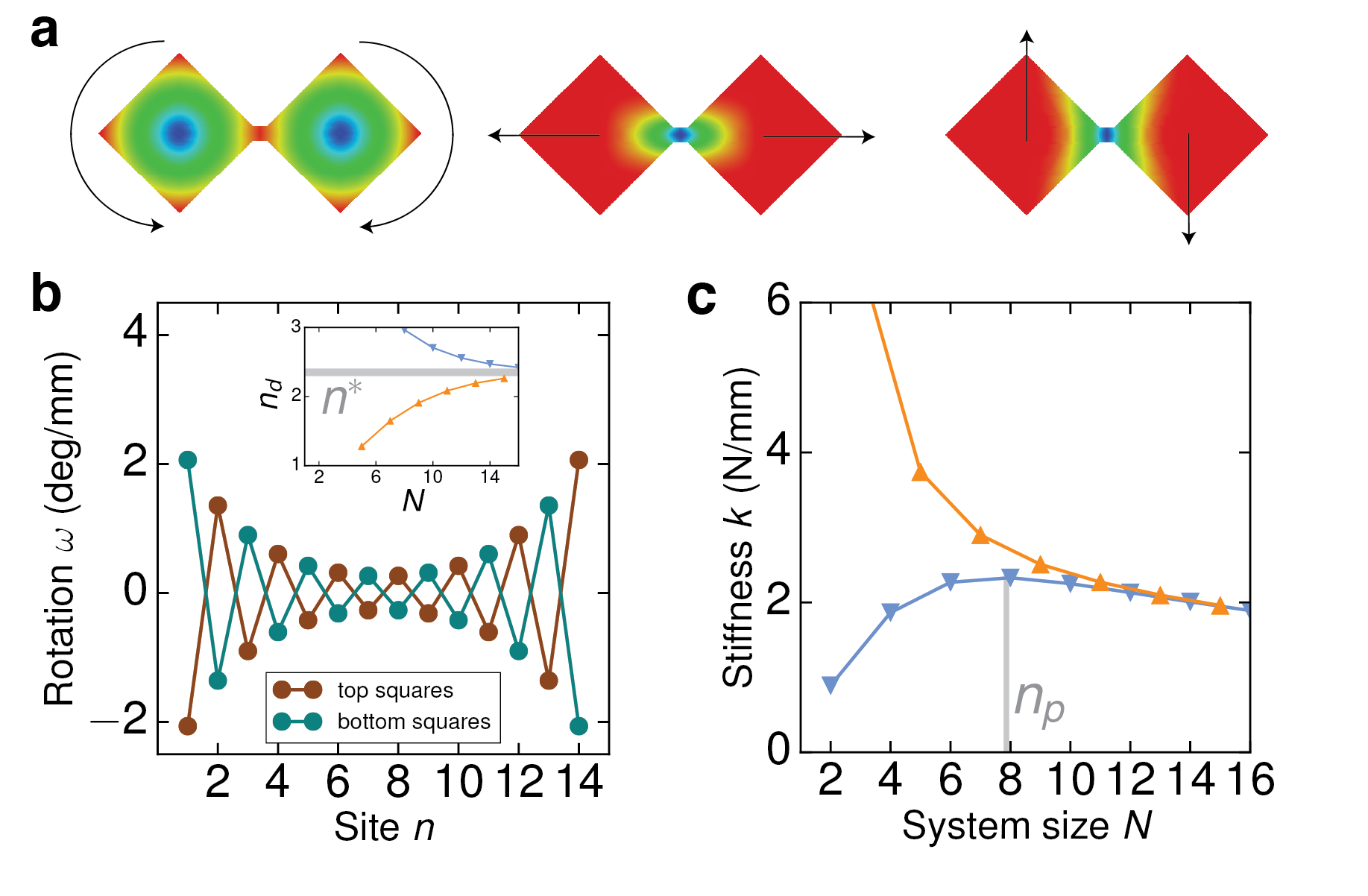}
\caption{{\bf Hybrid dressed mechanism.}
To check the applicability of the hybrid model (which we derive in detail in the Supplementary Information), we have determined the experimentally relevant values of $\alpha$ and $\beta$
using finite element simulations of the hinges, and solved the model using these numerical values.
(a) FEM determination of the bending, stretching and shear stiffness $C_b$, $k_j$ and $C_s$. The magnitude of the imposed displacements was $3\times10^{-4} ~L$, and the
color encodes the ratio of local over imposed displacements (blue: 0, red 1).
For the hinge parameters used here ($\ell/L = w/L = 0.1$), we find $C_b=1.62\times 10 ^1$~N.mm, $k_j=3.14\times 10 ^1$~N/mm, $C_s=1.82\times 10 ^1$~N.mm, leading to $\alpha=3.38\times 10^1$ and $\beta=1.39\times 10^2$. (b-c) Corresponding stiffnesses $k_o$ and $k_e$ and rotational rates $\omega$ are in excellent agreement with our experimental data.
{In particular, we find that the length scales $n^*=2.35$ and $n_p=6.9$ are in good agreement with the experimentally measured ones (displayed in Fig. 2 of the main text).}}
\label{ED:model}
\end{figure}

\clearpage
\onecolumngrid
\appendix
\section*{Supplementary Information}

\section{Mathematical description of the meta-chain}

The meta-chain described in the main text is based on the rotating squares mechanism~\cite{Grima_squares} and can be modelled with different degrees of complexity. In this supplementary document, we will adopt three modelling approaches. In all three of them, the squares are assumed infinitely rigid and are connected by flexible connections. In the first section, we assume that these connections can only bend, and in the second and third, we assume that they can also stretch and shear.
\subsection{Purely Rotating Square mechanism model}
We first describe the mechanical response of the mechanism depicted in Fig.~\ref{fig:SI0}, made of a periodic array of squares connected by their tips~\cite{Grima_squares} and actuated by a force $F$ at the outer most hinges. 
\subsubsection{Kinematics and geometrical constraints}
\begin{figure}[t!]
	\centering
	\includegraphics[width=0.6\columnwidth,trim=0cm 0cm 0cm 0cm]{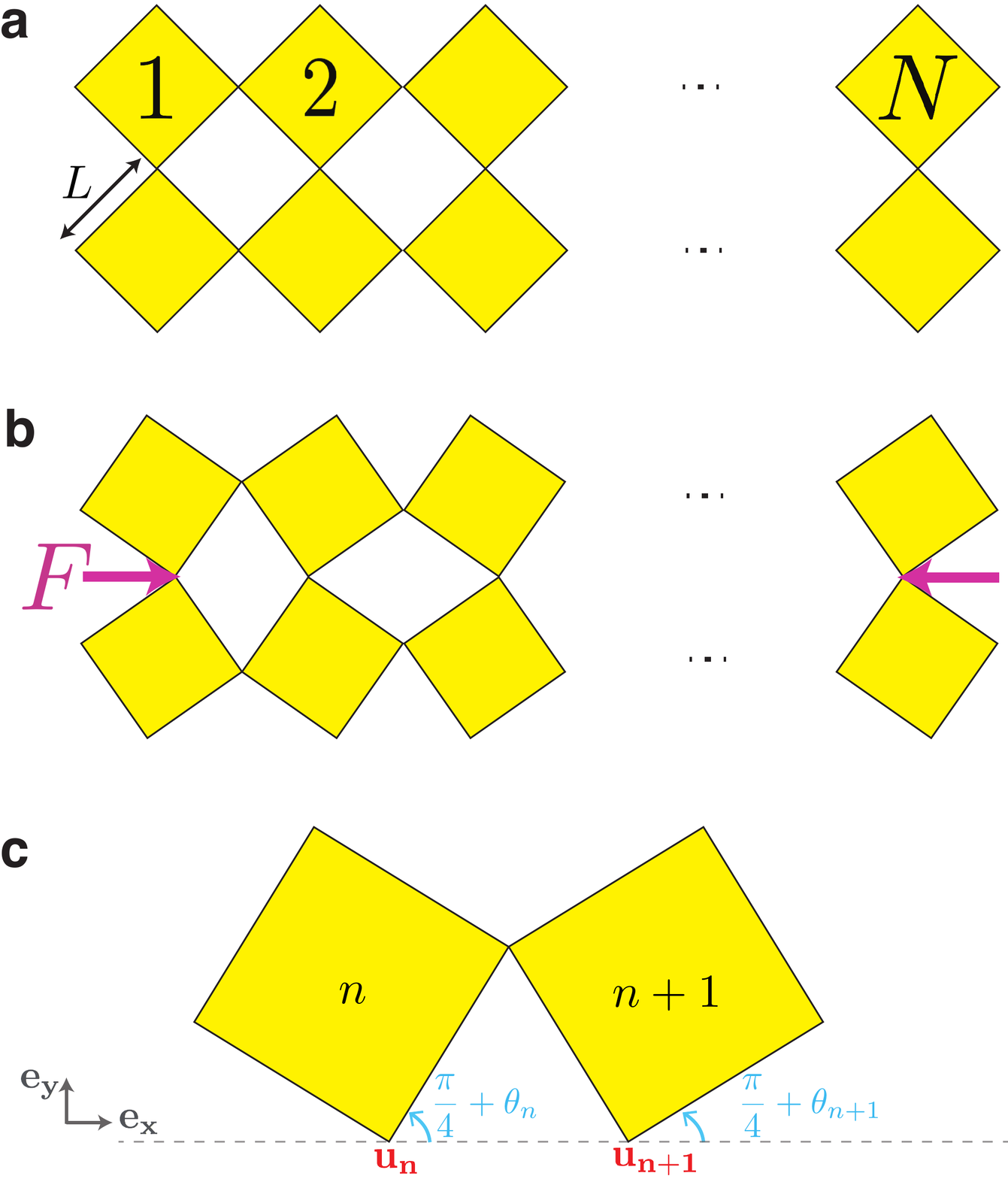}
	\caption{ab. The undeformed (a) and deformed (b) configurations of a rotating squares mechanism of length $N$. (c) A close-up of two neighbouring squares, showing the variables $\theta_n$ and $\theta_{n+1}=-\theta_n$.}
	\label{fig:SI0}
\end{figure}

The rotating squares mechanism allows only one mode of deformation, and the displacements of the central 
nodes $u_n$ (Fig.~\ref{fig:SI0}b) are related to the rotation of the squares as
\begin{equation}
\label{geometry0}
	u_{n+1} {-} u_n	= \frac{L}{\sqrt{2}} \left( \cos(\frac{\pi}{4} {+} \theta_n) 
	+ \cos(\frac{\pi}{4} {-} \theta_{n+1}) \right).
\end{equation}
As a result, the internal rotations and the end-to-end displacement can be expressed as follows
\begin{equation}
 u_N {-} u_1 = \frac{L}{\sqrt{2}}\sum_{n=1}^{N-1} \Big(  \cos(\frac{\pi}{4} {+} \theta_n) +  \cos(\frac{\pi}{4} {-} \theta_{n+1}) \Big).
\end{equation}
In addition, the rotation of subsequent squares are opposite (Fig.\ref{fig:SI0}b) and equal to $\theta_n=(-1)^n\Omega$. Therefore the above equation can be expressed as
\begin{equation}
u_N {-} u_1 =
\left\lbrace\begin{array}{ll} 
\frac{N-1}{2} \sqrt{2}L (\cos(\frac{\pi}{4} {+} \Omega)+\cos(\frac{\pi}{4} {-} \Omega))& \textrm{if $N$ is odd} \\[+3pt]
\frac{N-2}{2} \sqrt{2}L (\cos(\frac{\pi}{4} {+} \Omega)+\cos(\frac{\pi}{4} {-} \Omega))+\sqrt{2}L\cos(\frac{\pi}{4} {-} \Omega)& \textrm{if $N$ is even}\\
\end{array}\right.\label{eq:GE_Constraint}
\end{equation}

\subsubsection{Energetics and stiffness of the mechanism dressed with torsional springs}
Assuming that the bending of each hinge is penalised by an elastic energy, given by a torsional stiffness $C_b$, we can write
\begin{equation}
\label{eq:energy0}
E =  2\sum_{n=1}^{N-1} \left( \frac{C_b}{2} (\theta_n{-}\theta_{n+1})^2 \right)+ \sum_{n=1}^{N} \frac{C_b}{2}  (2\theta_n)^2,
\end{equation}
In addition, since the rotation of subsequent squares are opposite (Fig.\ref{fig:SI0}b), the elastic energy can be simply written as follows
\begin{equation}
\label{eq:energy0s}
E =  2(3N -2)C_b\,{\Omega}^2.
\end{equation}
Finally, in order to derive the stiffness, we assume that $\Omega\ll1$ and linearise Eq.~\ref{eq:GE_Constraint}
\begin{equation}
\delta=
\left\lbrace\begin{array}{ll} 
\mathcal{O}({\Omega}^2)& \textrm{if $N$ is odd} \\[+3pt]
L\Omega& \textrm{if $N$ is even}\\
\end{array}\right.,
\end{equation}
where $\delta=u_N {-} u_1 - (N-1)L$. This equation, combined with the identity $E=\tfrac{1}{2}k \delta^2$, yields
\begin{equation}
k=\left\lbrace\begin{array}{ll} \infty \,\, \qquad \qquad \qquad\textrm{if $N$ is odd} \\[+3pt]
4(3N-2)\frac{C_b}{L^2} \qquad \textrm{if $N$ is even}\\
\end{array}\right.
\end{equation}

\subsection{Hybrid mechanism model for the meta-chain}

\subsubsection{Kinematics and geometrical constraints}
In this model, we employ elastic hinges of length $\ell$ (Fig.~\ref{fig:SI1}a-c), with energy penalties associated to bending, stretch and shear (Fig.~\ref{fig:SI2}a-c). 
As the top and bottom rows behave symmetrically, we can describe the kinematics of the system by solely considering the top row (Fig.~\ref{fig:SI2}c). We describe the state of the system with the following variables: $\theta_n$ is the rotation of square $n$ compared to its starting configuration, $\psi_n$ is the angle that joint $n$ makes with the $x$-axis and $\varepsilon_n$ is the strain induced on joint $n$ (See Fig.~\ref{fig:SI1}c). We can express geometrically the distance between two subsequent bottom vertices of the squares $\vec{u}_n$, $\vec{u}_{n+1}$ as a function of the variables $\theta_n$, $\psi_n$, $\varepsilon_n$:

\begin{figure}[t!]
	\centering
	\includegraphics[width=0.6\columnwidth,trim= 0cm 0cm 0cm 0cm]{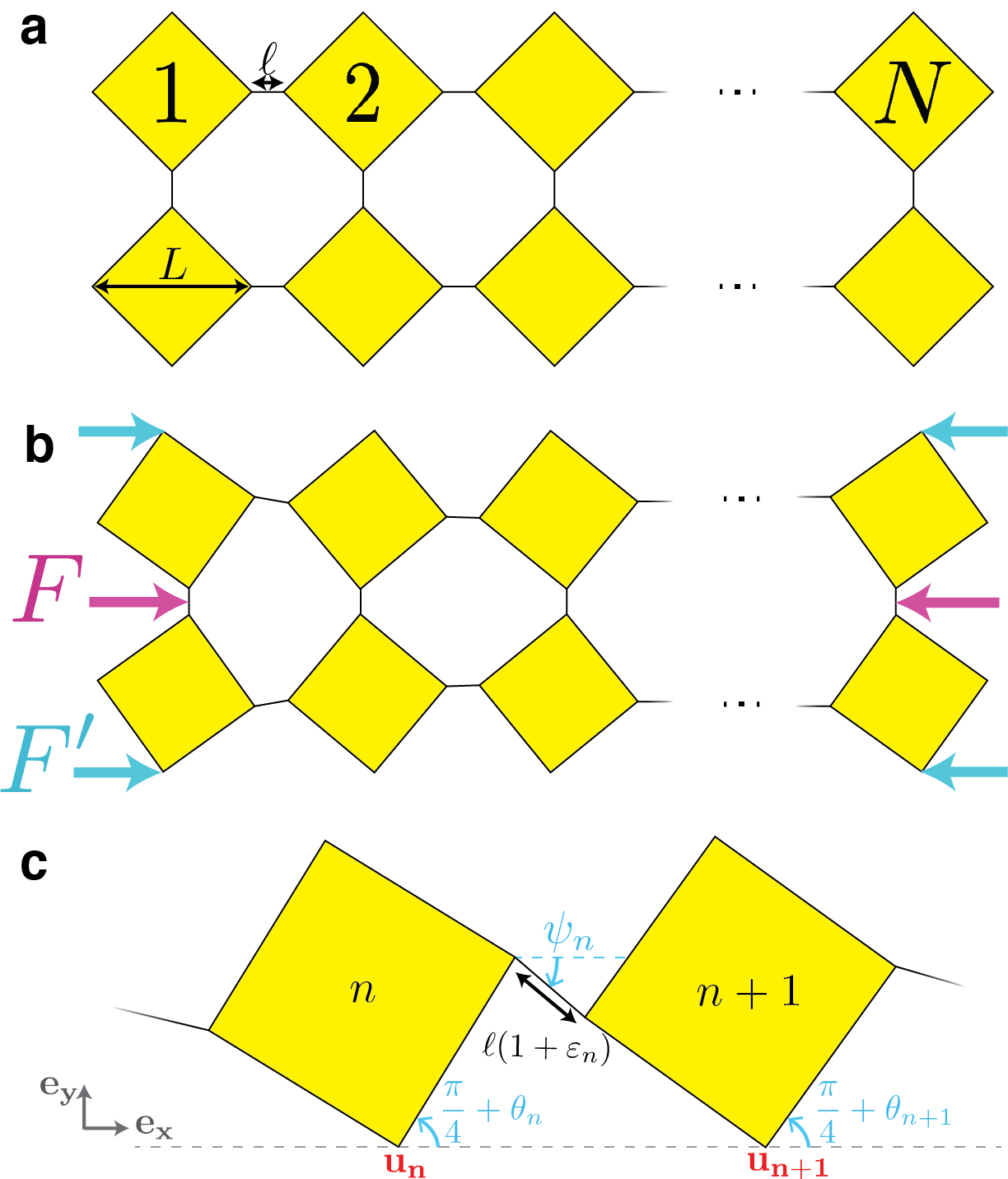}
	\caption{ab. The undeformed (a) and undeformed (b) configurations of a system of length $N$, showing the parameters $L$ and $\ell$. Our model comes in two flavours, where (i) the additional force $F'$ is zero and where (ii) $F'$ is finite.
	(b) A close-up of two neighbouring squares, showing the variables $\theta_n$, $\psi_n$, $\varepsilon_n$. The dotted line is the $x$-axis of the coordinate system.}
	\label{fig:SI1}
\end{figure}
\begin{equation} \label{geometry}
	\mathbf{u}_{n+1} {-} \mathbf{u}_n	= \frac{L}{\sqrt{2}} \begin{bmatrix} \cos(\sfrac{\pi}{4} {+} \theta_n) + \cos(\sfrac{\pi}{4} {-} \theta_{n+1}) \\ \sin(\sfrac{\pi}{4} {+} \theta_n) - \sin(\sfrac{\pi}{4} {-} \theta_{n+1}) \end{bmatrix} 
	+ \ell (1{+}\varepsilon_n) \begin{bmatrix} \cos(\psi_n) \\ \sin(\psi_n) \end{bmatrix}
\end{equation}
By symmetry, the square bottom vertices lie on the $x$-axis of the coordinate system. This condition translates into the constraints $c_1(n)=0$, for $ n \in [1,N{-}1]$ and $c_2=0$ with 
\begin{eqnarray}
	c_1(n)& =& \frac{L}{\sqrt{2}} \sin(\sfrac{\pi}{4} {+} \theta_n) - \frac{L}{\sqrt{2}} \sin(\sfrac{\pi}{4} {-} \theta_{n+1}) + \ell (1{+}\varepsilon_n) \sin(\psi_n)  \\
	c_2 & =& u_N {-} u_1 - \sum_{n=1}^{N-1} \Big( \frac{L}{\sqrt{2}} \cos(\sfrac{\pi}{4} {+} \theta_n) + \frac{L}{\sqrt{2}} \cos(\sfrac{\pi}{4} {-} \theta_{n+1}) + \ell (1{+}\varepsilon_n) \cos(\psi_n) \Big) .\label{eq:constraints}
\end{eqnarray}
The first constraint ensures that all points $\vec{u}_n$ lie on the $x$-axis, the second constraint connects the end-to-end distance of the system to the internal variables.

\subsubsection{Energetics and governing equations of the elastically dressed mechanism}

\begin{figure}[t!]
	\centering
\includegraphics[width=0.9\columnwidth,trim=0cm 9cm 0cm 7cm]{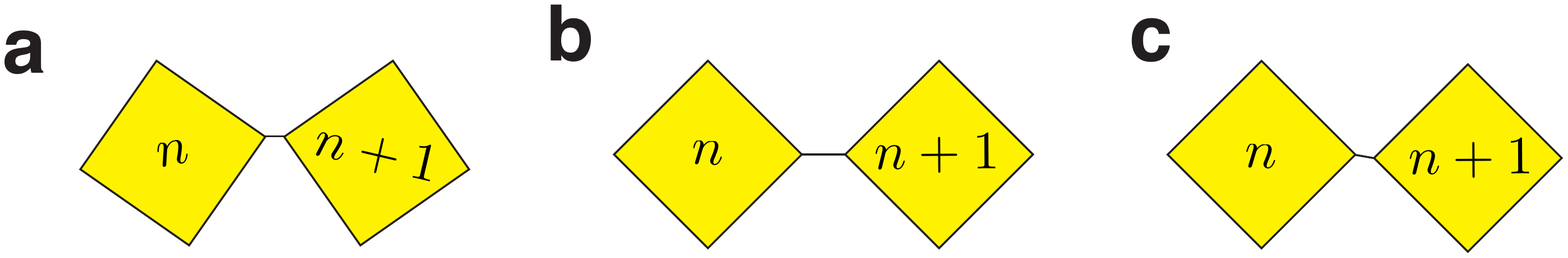}		
	\caption{The three deformation modes of the hybrid mechanism. (a) Pure bending, where $\theta_n -\theta_{n+1}\neq0$, $\psi_n = (\theta_n + \theta_{n+1})/2$ and $\varepsilon_n=0$. (b) Pure stretch, where $\varepsilon_n\neq0$, $\theta_n = \theta_{n+1}$ and $\psi_n = (\theta_n + \theta_{n+1})/2$.
	(c) Pure shear, where $\psi_n \neq (\theta_n + \theta_{n+1})/2$, $\theta_n = \theta_{n+1}$ and $\varepsilon_n=0$.}
	\label{fig:SI2}
\end{figure}

This structure has multiple degrees of freedom, which when actuated, cost elastic energy. We assume that (a) the pure bending of each connection is governed by the torsional stiffness $C_b$ (Fig.~\ref{fig:SI2}a); (b) the stretching of each connections is governed by the linear stiffness $k_j$ (Fig.~\ref{fig:SI2}b); (c) the pure shear of each connection is governed by the torsional stiffness $C_s$ (Fig.~\ref{fig:SI2}c). One should not be surprised by the fact that $C_b$ and $C_s$ are a priori different. In a fully elastic structure (See e.g. Extended Data Figs.~3,4 of the main text),  the hinge bending and shear are associated to the same type of local deformations, yet are localised at different places within the filament that acts as a hinge. Combining these stiffnesses to the kinematics expressed above, we can then express the elastic energy of the system as follows:

\begin{equation} \label{energy}
E =  2\sum_{n=1}^{N-1} \left( \frac{C_b}{2} (\theta_n{-}\theta_{n+1})^2 + \frac{C_s}{2} \left( \frac{\theta_n {+} \theta_{n+1}}{2} {-} \psi_n \right)^2 + \frac{k_j}{2}  (\ell \varepsilon_n)^2 \right)  + \sum_{n=1}^{N} \frac{C_b}{2}  (2\theta_n)^2,
\end{equation}
where the first sum corresponds to the energy of the two rows of horizontal connections, which can both bend, stretch and shear and the second sum to the vertical connections, which by symmetry only experience bending.

\subsubsection{Equations governing the mechanical equilibrium}

The elastic energy expressed in Eq. (\ref{energy}) has to be minimized in the presence of the geometrical constraints $c_1(n)=0$ and $c_2=0$. To do this,  we introduce the Lagrange function:

\begin{equation} \label{lagrangian}
\begin{aligned}
	\mathcal{L} &= E - \sum_{n=1}^{N-1} G_n c_1(n) - F c_2.
\end{aligned}
\end{equation}
Here $G_n$, $F$ are the Lagrange multipliers, where $F$ corresponds to the force applied in the $x$-direction. Mechanical equilibria are found at stationary points of the Lagrange function, which are given when the partial derivatives of the Lagrange function with respect to the variables $\theta_n$, $\psi_n$, $\varepsilon_n$, $G_n$, $F$ are zero. Since we focus on solutions for small displacements, the equations found are subsequently linearised with respect to these variables. After a few algebraical manipulations and substitutions, we can express the governing equations solely with respect to the angle $\theta_n$ and the force $F$:
\begin{subequations} \label{solvet}
\begin{align}
\begin{split}
	\frac{C_s}{4C_b} \left( 1 {+} \frac{L}{\ell} \right)^2 (\theta_1 {+} \theta_2) &=  (\theta_2 {-} 3 \theta_1) +\frac{ L }{4 C_b} F\label{solvet0}
\end{split} \\
\begin{split}
	\frac{C_s}{4C_b}  \left( 1 {+} \frac{L}{\ell} \right)^2 (\theta_{n-1} {+} 2 \theta_n {+} \theta_{n+1}) &=  (\theta_{n-1} {-} 4 \theta_n {+} \theta_{n+1}) \label{solvetn}
\end{split} \quad\textrm{ for } n \in [2,N{-}1]\\
\begin{split}
	\frac{C_s}{4C_b}  \left( 1 {+} \frac{L}{\ell} \right)^2 (\theta_{N-1} {+} \theta_N) &=   (\theta_{N-1} {-} 3 \theta_N) - \frac{ L }{4 C_b} F \label{solvetN}
\end{split},
\end{align}
\end{subequations}
and
\begin{equation}
	 (N-1) F = - 2 k_j \delta + L k_j  (\theta_N {-} \theta_1), \label{eq:solveF}
\end{equation}
where $\delta \equiv u_N {-} u_1 - (N-1) ( L {+} \ell)$ is the structure's compressive displacement, i.e. is negative (positive) under compression (tension). 
Note that the stretch $\varepsilon_n=-F /2 k_j \ell$ is independent of the discrete coordinate $n$.
After a non-dimensionalization step, we find the equations
\begin{subequations} \label{solvet}
\begin{align}
\begin{split}
	\alpha( \theta_0 {+} \theta_1) &=  (\theta_1 {-} 3 \theta_0) + \tilde{F}\label{solvet0}
\end{split} \\
\begin{split}
	\alpha  (\theta_{n-1} {+} 2 \theta_n {+} \theta_{n+1}) &=  (\theta_{n-1} {-} 4 \theta_n {+} \theta_{n+1}) \label{solvetn}
\end{split} \quad\textrm{ for } n \in [2,N{-}1]\\
\begin{split}
	\alpha (\theta_{N-1} {+} \theta_N) &=   (\theta_{N-1} {-} 3 \theta_N) - \tilde{F}\label{solvetN}
\end{split}\\
\begin{split}(N-1) \beta^{-1} F &= (\theta_N {-} \theta_1) - \tilde{\delta}\end{split}.\label{solveF}
\end{align}
\end{subequations}
Here the non-dimensional parameters $\alpha=\tfrac{Cs}{4C_b}(1+\tfrac{L}{\ell})^2$ and $\beta=\tfrac{k_j L^2}{4C_b}$ represent the relative cost of shear and stretch to bending, respectively, and are the eqs. (1) of the main text. In addition, the non-dimensionalised force and displacement read respectively $\tilde{F}=F \tfrac{L}{4C_b}$ and $\tilde{\delta}=\tfrac{2\delta}{L}$. 

This system of equations is linear and can possibly be solved analytically for each value of $N$. However, the expressions differ for every value of $N$ and become impractically large for large $N$. Therefore, we solved the equations numerically for each value of $N$, $\alpha$ and $\beta$ in order to calculate the stiffness and rotational field (Fig. 3 of the main text and Extended Data Figure 4). The peak length scale $n_p$ is determined by using the location of the maximum of a quadratic fit to the numerically estimated stiffness $k_e$ vs. $N$ in the vicinity of the maximum value of $k_e$. The characteristic length scale $n^*$ is determined by using the decay length of an exponential fit to the rotational field of a meta-chain of length $N=1000$.

\subsubsection{Continuum limit of the bulk equation}
In order to obtain an explicit mathematical expression for the characteristic length $n^*$,
it is worthwhile to consider Eq.~(\ref{solvetn}) in the continuum limit. Assuming that the envelope of the counter-rotating field has small gradients, we can perform a Taylor expansion of the discrete staggered field around $x=n$, $\theta_n=\tilde{\theta}(x)$, $\theta_{n+1}= - \tilde{\theta}(x) - b\tilde{\theta}_x(x) - \tfrac{b^2}{2} \tilde{\theta}_{xx}(x)$ and $\theta_{n-1}= - \tilde{\theta}(x)  + b\tilde{\theta}_x(x) - \tfrac{b^2}{2} \tilde{\theta}_{xx}(x)$, where $b= L {+} \ell$ is the distance between two squares. Eqs.~(\ref{solvetn}) then becomes
\begin{equation}
 \frac{b^2}{6} (\alpha-1) \tilde{\theta}_{xx} - \tilde{\theta}=0.
\end{equation}
As a result, the family of solutions for this continuum staggered field $\tilde{\theta}$ 
is $\lbrace \exp x/(b n^*),\exp - x/(b n^*)\rbrace$, where $n^*= \sqrt{\tfrac{1}{6} (\alpha-1)}$, whose scaling in the large $\alpha$ limit is consistent with the numerical solution of Eqs.~(\ref{solvet0}-\ref{solveF}) discussed in the main text. This continuum approach establishes a clear link between the intrinsic lengthscale $n^*$ and the decay length that appears in the presence of boundaries or inhomogeneities.

\subsection{Hybrid mechanism model for the meta-chain with complex boundary conditions}
In this section we probe the mechanical response of the meta-chain by applying a load $F'$ at additional most upper and lower vertices of the edge squares (See Fig.~\ref{fig:SI1}b). As a result, the end-to-end vertices undergo a relative horizontal displacement $\delta'$. This additional loading condition translates as an additional constraint, written as follows
\begin{equation}
	c_3  = u_N' {-} u_1' - \sum_{n=1}^{N-1} \Big( \frac{L}{\sqrt{2}} \cos(\sfrac{\pi}{4} {+} \theta_n) + \frac{L}{\sqrt{2}} \cos(\sfrac{\pi}{4} {-} \theta_{n+1}) + \ell (1{+}\varepsilon_n) \cos(\psi_n) \Big) -L\sin(\theta_1)-L\sin(\theta_N).\label{eq:constraintsb}
\end{equation}
This additional constraint leads to
\begin{subequations} \label{Hsolvet}
\begin{align}
\begin{split}
	\alpha( \theta_1 {+} \theta_2) &=  (\theta_2 {-} 3 \theta_1) + (\tilde{F}-2\tilde{F'})\label{Hsolvet0}
\end{split} \\
\begin{split}
	\alpha  (\theta_{n-1} {+} 2 \theta_n {+} \theta_{n+1}) &=  (\theta_{n-1} {-} 4 \theta_n {+} \theta_{n+1}) \label{Hsolvetn}
\end{split} \quad\textrm{ for } n \in [2,N{-}1]\\
\begin{split}
	\alpha (\theta_{N-1} {+} \theta_N) &=   (\theta_{N-1} {-} 3 \theta_N) - (\tilde{F}-2\tilde{F'})\label{HsolvetN}
\end{split},
\end{align}
\end{subequations}
and
\begin{eqnarray}
	 (N-1) \beta^{-1} (\tilde{F}+2\tilde{F'}) &=& (\theta_N {-} \theta_1) - \tilde{\delta} , \label{eq:solveF}\\
	 (N-1) \beta^{-1} (\tilde{F}+2\tilde{F'})	 &=& -(\theta_N {-} \theta_1) - \tilde{\delta'} ,
\end{eqnarray}
where $2\tilde{F'}$ corresponds to the total force applied at the outer---top and bottom---vertices of the chain. Note that the stretch $\varepsilon_n=-(\tilde{F}+2\tilde{F'}) /2 k_j \ell$ is independent of the discrete coordinate $n$. We calculate the effective stiffness of the structure as follows
\begin{equation}
\tilde{k}=\frac{\tilde{F}}{\tilde{\delta}}+2\frac{\tilde{F'}}{\tilde{\delta'}}.
\end{equation}

\section{Mathematical description of the topological metamaterial}

\begin{figure}[t!]
	\centering
\includegraphics[width=0.9\columnwidth,trim=0cm 0cm 0cm 0cm]{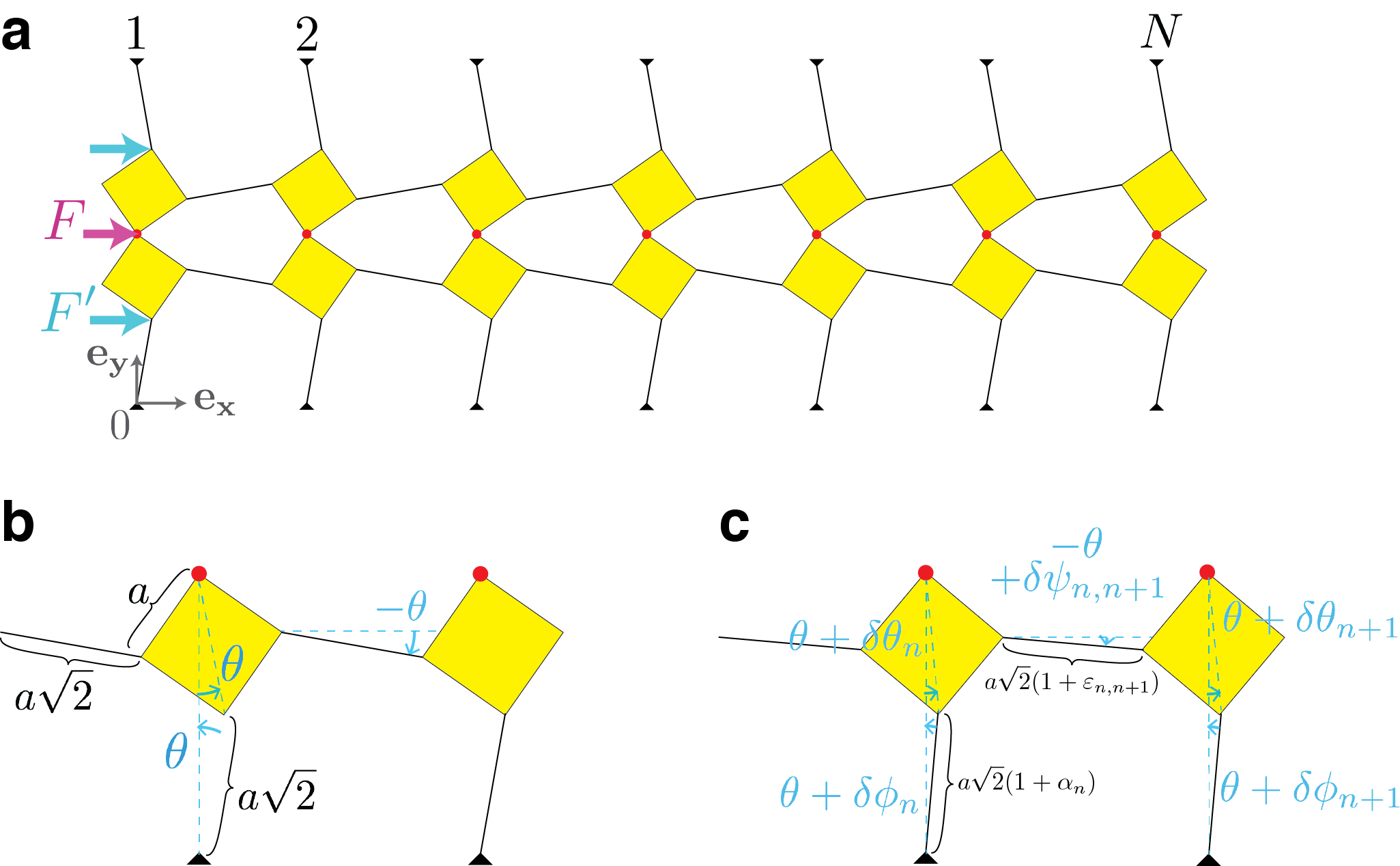}		
	\caption{Sketch of the mechanical metamaterial 2. a. Geometry of the chain constituted of $N$ unit cells, characterized by their initial tilt angle $\theta$ and connected in the middle by a torsional spring $C$ (red dots). bc. Geometry of two unit cells in undeformed (b) and deformed (c) configurations. For simplificity, in the text we work the condition $a=1$, without loss of generality. Adapted from ~\cite{Coulais_nonreci}.}
	\label{fig:SI4}
\end{figure}

In Figure 5 of the main text, we introduce a topological mechanical metamaterial (See Fig.~\ref{fig:SI4}a) which was previously studied by Coulais et al.~\cite{Coulais_nonreci}. By contrast with the previous study where only the pure mechanism was considered under simple boundary conditions, we here focus on the mechanical response of the mechanism-based metamaterial by taking into account the elastic hinge deformations which compete with the mechanism and by using more generic boundary conditions. The system Lagrangian reads
\begin{equation}
\begin{split}\mathcal{L}=&
\frac{1}{2} C \sum _{n=1}^N \delta \theta _n^2+\frac{1}{2} k \sum _{n=1}^N \alpha _n^2+\frac{1}{2} k \sum _{n=1}^{N-1} \varepsilon _n^2\\
&+
	\begin{split}&\sum _{n=1}^{N-1} 
		\kappa _n \left(2 \sqrt{2} \cos (\theta )+\sqrt{2} \left(\left(\alpha _n+1\right) 					\left(-\sin \left(\theta +\delta \phi _n\right)\right)+\left(\alpha _{n+1}+1\right) 					\sin \left(\theta +\delta \phi _{n+1}\right)\right.\right.
		\\&\left.\left.
		-\left(\varepsilon _{n,n+1}+1\right) \cos \left(\theta -\delta \psi _{n,n+1}\right)\right)-\sin \left(-				\theta -\delta \theta _n+\frac{\pi }{4}\right)-\sin \left(\theta +\delta \theta _{n+1}+				\frac{\pi }{4}\right)\right)
	\end{split}
\\&+
	\begin{split}&\sum _{n=1}^{N-1} \lambda _n \left(\sqrt{2} \left(\left(\alpha _n+1\right) \left(-\cos \left(\theta +\delta \phi _n\right)\right)+\left(\alpha _{n+1}+1\right) \cos \left(\theta +\delta \phi _{n+1}\right)\right.\right.
\\&	\left.\left.
	+\left(\varepsilon _{n,n+1}+1\right) \sin \left(\theta -\delta \psi _{n,n+1}\right)\right)-\cos \left(-\theta -\delta \theta _n+\frac{\pi }{4}\right)+\cos \left(\theta +\delta \theta _{n+1}+\frac{\pi }{4}\right)\right)
	\end{split}
\\&
+\sum _{n=1}^N \mu _n \left(-2 \cos (\theta )+\left(\alpha _n+1\right) \cos \left(\theta +\delta \phi _n\right)+\cos \left(\theta +\delta \theta _n\right)\right)
\\&+\sqrt{2} F \left(\left(\alpha _1+1\right) \sin \left(\delta \phi _1+\theta \right)-\sin \left(\delta \theta _1+\theta \right)\right)+2\sqrt{2}F' \left(\alpha _1+1\right)  \sin \left(\delta \phi _1+\theta \right)
\end{split},
\end{equation}
where the quantities $\delta\theta_i$, $\alpha_i$, $\varepsilon_{i,i+1}$ and $\delta\psi_{i,i+1}$ are internal degrees of freedom of the structure. The quantities $\lambda_i$, $\mu_i$, $\kappa_i$, $F$ and $F'$ are Lagrange multipliers associated to the geometric constraints.

Mechanical equilibria are found at the stationary points of the Lagrangian, therefore follow from the equations $\partial \mathcal{L}/\partial \delta\theta_i = 0$, $\partial \mathcal{L}/\partial \delta\alpha_i=0$, and so on. After a few simple algebraic manipulations and substitutions, we find the following equations
\begin{subequations} \label{TOPOHsolvet}
\begin{align}
\begin{split}
-2 \sqrt{2} \frac{2 F+2F'}{k} \sin (2 \theta )=&
\sin (\theta ) \left(-4 \frac{C}{k} \delta \theta _1 +\left(4 c_1 c_2\delta \theta _2 -4 c_1^2\delta \theta _1 \right)\right)\\&+\left(\alpha _1-\alpha _2\right) (\sin\theta+\sin (3 \theta )+5 \cos (\theta )+\cos (3 \theta ))
\end{split}
\\ 
\begin{split}0=&
4\frac{C}{k} \delta \theta_n-4 c_1 c_2\delta (\theta _{n-1}+\delta \theta _{n+1}) + 4 (c_1^2 + c_2^2)\delta \theta _n  \\ &
+ 2( (\cot \theta +1)(c_1+c_2)-1)\alpha _{n+1} - 2(2c_1-3\cot\theta-2)\alpha _{n-1}-4 \cot (\theta )(c_1+c_2) \alpha _n,
\end{split}
\\
\begin{split}0=&4 \frac{C}{k} \delta \theta_{N}-4c_1c_2\delta \theta _{N-1} +4c_2^2\delta \theta _N+4c_2\cot (\theta )  \left(\alpha _{N-1}-\alpha _N\right) 
\end{split}
\end{align}
\end{subequations}
and
\begin{subequations}
\begin{align}
\begin{split}-
2 \sqrt{2} \frac{F+2F'}{k} \csc\theta=&
 2\cot \theta  \left(- c_1\delta \theta _1 +c_2\delta \theta _2 \right) +2\csc ^2\theta  \alpha _1-2 \alpha _2 \cot^2 \theta  
\end{split}
\\
\begin{split}0=&
2\cot \theta  \left(c_1 \delta\theta_{n-1} -2 (c_1+c_2)\delta\theta_{n} + c_2 \delta\theta_{n+1}\right)+2(1+2\cot ^2\theta )\alpha _n -2\cot ^2\theta \left(\alpha _{n-1}+\alpha _{n+1}\right)
\end{split}\\
\begin{split}0=&
2\cot (\theta ) \left(c_1\delta \theta _{N-1} -c_2\delta \theta _N \right)+2 \alpha _N \csc ^2(\theta )-2 \alpha _{N-1} \cot ^2(\theta ),
\end{split}
\end{align}
\end{subequations}
with $c_1=\frac{1}{2} (\sin (2 \theta )+\cos (2 \theta )+2)$ and $c_2=\frac{1}{2} (-\sin (2 \theta )+\cos (2 \theta )+2)$. To produce the results shown in figure 5d-f of the main text, 
we solve these equations numerically for $C=0.1$, $k=10$,  $\theta=\pi/16$ and varying the ratio between $F$ and $F'$. The hybridisation of the left-localised and right-localised deformation modes occurs because the deformation fields $\alpha_n$ and $\delta\theta_n$ are mixed.


\begin{thebibliography}{29}%
\makeatletter
\providecommand \@ifxundefined [1]{%
 \@ifx{#1\undefined}
}%
\providecommand \@ifnum [1]{%
 \ifnum #1\expandafter \@firstoftwo
 \else \expandafter \@secondoftwo
 \fi
}%
\providecommand \@ifx [1]{%
 \ifx #1\expandafter \@firstoftwo
 \else \expandafter \@secondoftwo
 \fi
}%
\providecommand \natexlab [1]{#1}%
\providecommand \enquote  [1]{``#1''}%
\providecommand \bibnamefont  [1]{#1}%
\providecommand \bibfnamefont [1]{#1}%
\providecommand \citenamefont [1]{#1}%
\providecommand \href@noop [0]{\@secondoftwo}%
\providecommand \href [0]{\begingroup \@sanitize@url \@href}%
\providecommand \@href[1]{\@@startlink{#1}\@@href}%
\providecommand \@@href[1]{\endgroup#1\@@endlink}%
\providecommand \@sanitize@url [0]{\catcode `\\12\catcode `\$12\catcode
  `\&12\catcode `\#12\catcode `\^12\catcode `\_12\catcode `\%12\relax}%
\providecommand \@@startlink[1]{}%
\providecommand \@@endlink[0]{}%
\providecommand \url  [0]{\begingroup\@sanitize@url \@url }%
\providecommand \@url [1]{\endgroup\@href {#1}{\urlprefix }}%
\providecommand \urlprefix  [0]{URL }%
\providecommand \Eprint [0]{\href }%
\providecommand \doibase [0]{http://dx.doi.org/}%
\providecommand \selectlanguage [0]{\@gobble}%
\providecommand \bibinfo  [0]{\@secondoftwo}%
\providecommand \bibfield  [0]{\@secondoftwo}%
\providecommand \translation [1]{[#1]}%
\providecommand \BibitemOpen [0]{}%
\providecommand \bibitemStop [0]{}%
\providecommand \bibitemNoStop [0]{.\EOS\space}%
\providecommand \EOS [0]{\spacefactor3000\relax}%
\providecommand \BibitemShut  [1]{\csname bibitem#1\endcsname}%
\let\auto@bib@innerbib\@empty
\bibitem [{\citenamefont {Lakes}(1987)}]{Lakes_poisson}%
  \BibitemOpen
  \bibfield  {author} {\bibinfo {author} {\bibfnamefont {R.}~\bibnamefont
  {Lakes}},\ }\bibfield  {title} {\enquote {\bibinfo {title} {Foam structures
  with a negative poisson's ratio},}\ }\href
  {http://www.jstor.org/stable/1698767} {\bibfield  {journal} {\bibinfo
  {journal} {Science}\ }\textbf {\bibinfo {volume} {235}},\ \bibinfo {pages}
  {1038--1040} (\bibinfo {year} {1987})}\BibitemShut {NoStop}%
\bibitem [{\citenamefont {Grima}\ and\ \citenamefont
  {Evans}(2000)}]{Grima_squares}%
  \BibitemOpen
  \bibfield  {author} {\bibinfo {author} {\bibfnamefont {J.~N.}\ \bibnamefont
  {Grima}}\ and\ \bibinfo {author} {\bibfnamefont {K.~E.}\ \bibnamefont
  {Evans}},\ }\bibfield  {title} {\enquote {\bibinfo {title} {Auxetic behavior
  from rotating squares},}\ }\href {\doibase Doi 10.1023/A:1006781224002}
  {\bibfield  {journal} {\bibinfo  {journal} {J. Mater. Sc. Lett.}\ }\textbf
  {\bibinfo {volume} {19}},\ \bibinfo {pages} {1563--1565} (\bibinfo {year}
  {2000})}\BibitemShut {NoStop}%
\bibitem [{\citenamefont {Kadic}\ \emph {et~al.}(2012)\citenamefont {Kadic},
  \citenamefont {B\"uckmann}, \citenamefont {Stenger}, \citenamefont {Thiel},\
  and\ \citenamefont {Wegener}}]{Kadic_pentamode}%
  \BibitemOpen
  \bibfield  {author} {\bibinfo {author} {\bibfnamefont {M.}~\bibnamefont
  {Kadic}}, \bibinfo {author} {\bibfnamefont {T.}~\bibnamefont {B\"uckmann}},
  \bibinfo {author} {\bibfnamefont {N.}~\bibnamefont {Stenger}}, \bibinfo
  {author} {\bibfnamefont {M.}~\bibnamefont {Thiel}}, \ and\ \bibinfo {author}
  {\bibfnamefont {M.}~\bibnamefont {Wegener}},\ }\bibfield  {title} {\enquote
  {\bibinfo {title} {On the practicability of pentamode mechanical
  metamaterials},}\ }\href
  {http://scitation.aip.org/content/aip/journal/apl/100/19/10.1063/1.4709436}
  {\bibfield  {journal} {\bibinfo  {journal} {Appl. Phys. Lett.}\ }\textbf
  {\bibinfo {volume} {100}},\ \bibinfo {pages} {191901} (\bibinfo {year}
  {2012})}\BibitemShut {NoStop}%
\bibitem [{\citenamefont {Wei}\ \emph {et~al.}(2013)\citenamefont {Wei},
  \citenamefont {Guo}, \citenamefont {Dudte}, \citenamefont {Liang},\ and\
  \citenamefont {Mahadevan}}]{Maha_miura}%
  \BibitemOpen
  \bibfield  {author} {\bibinfo {author} {\bibfnamefont {Z.~Y.}\ \bibnamefont
  {Wei}}, \bibinfo {author} {\bibfnamefont {Z.~V.}\ \bibnamefont {Guo}},
  \bibinfo {author} {\bibfnamefont {L.}~\bibnamefont {Dudte}}, \bibinfo
  {author} {\bibfnamefont {H.~Y.}\ \bibnamefont {Liang}}, \ and\ \bibinfo
  {author} {\bibfnamefont {L.}~\bibnamefont {Mahadevan}},\ }\bibfield  {title}
  {\enquote {\bibinfo {title} {Geometric mechanics of periodic pleated
  origami},}\ }\href {\doibase 10.1103/PhysRevLett.110.215501} {\bibfield
  {journal} {\bibinfo  {journal} {Phys. Rev. Lett.}\ }\textbf {\bibinfo
  {volume} {110}},\ \bibinfo {pages} {215501} (\bibinfo {year}
  {2013})}\BibitemShut {NoStop}%
\bibitem [{\citenamefont {Bückmann}\ \emph
  {et~al.}(2014{\natexlab{a}})\citenamefont {Bückmann}, \citenamefont
  {Schittny}, \citenamefont {Thiel}, \citenamefont {Kadic}, \citenamefont
  {Milton},\ and\ \citenamefont {Wegener}}]{Wegener_Milton}%
  \BibitemOpen
  \bibfield  {author} {\bibinfo {author} {\bibfnamefont {T.}~\bibnamefont
  {Bückmann}}, \bibinfo {author} {\bibfnamefont {R.}~\bibnamefont {Schittny}},
  \bibinfo {author} {\bibfnamefont {M.}~\bibnamefont {Thiel}}, \bibinfo
  {author} {\bibfnamefont {M.}~\bibnamefont {Kadic}}, \bibinfo {author}
  {\bibfnamefont {G.~W.}\ \bibnamefont {Milton}}, \ and\ \bibinfo {author}
  {\bibfnamefont {M.}~\bibnamefont {Wegener}},\ }\bibfield  {title} {\enquote
  {\bibinfo {title} {On three-dimensional dilational elastic metamaterials},}\
  }\href {http://stacks.iop.org/1367-2630/16/i=3/a=033032} {\bibfield
  {journal} {\bibinfo  {journal} {New J. Phys.}\ }\textbf {\bibinfo {volume}
  {16}},\ \bibinfo {pages} {033032} (\bibinfo {year}
  {2014}{\natexlab{a}})}\BibitemShut {NoStop}%
\bibitem [{\citenamefont {Bückmann}\ \emph
  {et~al.}(2014{\natexlab{b}})\citenamefont {Bückmann}, \citenamefont {Thiel},
  \citenamefont {Kadic}, \citenamefont {Schittny},\ and\ \citenamefont
  {Wegener}}]{Wegener_cloak}%
  \BibitemOpen
  \bibfield  {author} {\bibinfo {author} {\bibfnamefont {T.}~\bibnamefont
  {Bückmann}}, \bibinfo {author} {\bibfnamefont {M.}~\bibnamefont {Thiel}},
  \bibinfo {author} {\bibfnamefont {M.}~\bibnamefont {Kadic}}, \bibinfo
  {author} {\bibfnamefont {R.}~\bibnamefont {Schittny}}, \ and\ \bibinfo
  {author} {\bibfnamefont {M.}~\bibnamefont {Wegener}},\ }\bibfield  {title}
  {\enquote {\bibinfo {title} {An elasto-mechanical unfeelability cloak made of
  pentamode metamaterials},}\ }\href {http://dx.doi.org/10.1038/ncomms5130}
  {\bibfield  {journal} {\bibinfo  {journal} {Nat. Commun.}\ }\textbf {\bibinfo
  {volume} {5}} (\bibinfo {year} {2014}{\natexlab{b}})}\BibitemShut {NoStop}%
\bibitem [{\citenamefont {Mullin}\ \emph {et~al.}(2007)\citenamefont {Mullin},
  \citenamefont {Deschanel}, \citenamefont {Bertoldi},\ and\ \citenamefont
  {Boyce}}]{Mullin_holey}%
  \BibitemOpen
  \bibfield  {author} {\bibinfo {author} {\bibfnamefont {T.}~\bibnamefont
  {Mullin}}, \bibinfo {author} {\bibfnamefont {S.}~\bibnamefont {Deschanel}},
  \bibinfo {author} {\bibfnamefont {K.}~\bibnamefont {Bertoldi}}, \ and\
  \bibinfo {author} {\bibfnamefont {M.~C.}\ \bibnamefont {Boyce}},\ }\bibfield
  {title} {\enquote {\bibinfo {title} {Pattern transformation triggered by
  deformation},}\ }\href
  {http://link.aps.org/doi/10.1103/PhysRevLett.99.084301} {\bibfield  {journal}
  {\bibinfo  {journal} {Phys. Rev. Lett.}\ }\textbf {\bibinfo {volume} {99}},\
  \bibinfo {pages} {084301} (\bibinfo {year} {2007})}\BibitemShut {NoStop}%
\bibitem [{\citenamefont {Nicolaou}\ and\ \citenamefont
  {Motter}(2012)}]{Motter_negcompress}%
  \BibitemOpen
  \bibfield  {author} {\bibinfo {author} {\bibfnamefont {Z.~G.}\ \bibnamefont
  {Nicolaou}}\ and\ \bibinfo {author} {\bibfnamefont {A.~E.}\ \bibnamefont
  {Motter}},\ }\bibfield  {title} {\enquote {\bibinfo {title} {Mechanical
  metamaterials with negative compressibility transitions},}\ }\href
  {http://dx.doi.org/10.1038/nmat3331} {\bibfield  {journal} {\bibinfo
  {journal} {Nat. Mater.}\ }\textbf {\bibinfo {volume} {11}},\ \bibinfo {pages}
  {608--613} (\bibinfo {year} {2012})}\BibitemShut {NoStop}%
\bibitem [{\citenamefont {Shim}\ \emph {et~al.}(2012)\citenamefont {Shim},
  \citenamefont {Perdigou}, \citenamefont {Chen}, \citenamefont {Bertoldi},\
  and\ \citenamefont {Reis}}]{Shim_buckliball}%
  \BibitemOpen
  \bibfield  {author} {\bibinfo {author} {\bibfnamefont {J.}~\bibnamefont
  {Shim}}, \bibinfo {author} {\bibfnamefont {C.}~\bibnamefont {Perdigou}},
  \bibinfo {author} {\bibfnamefont {E.~R.}\ \bibnamefont {Chen}}, \bibinfo
  {author} {\bibfnamefont {K.}~\bibnamefont {Bertoldi}}, \ and\ \bibinfo
  {author} {\bibfnamefont {P.~M.}\ \bibnamefont {Reis}},\ }\bibfield  {title}
  {\enquote {\bibinfo {title} {Buckling-induced encapsulation of structured
  elastic shells under pressure},}\ }\href
  {http://www.pnas.org/content/109/16/5978.abstract} {\bibfield  {journal}
  {\bibinfo  {journal} {Proc. Natl. Ac. Sc. U. S. A.}\ }\textbf {\bibinfo
  {volume} {109}},\ \bibinfo {pages} {5978--5983} (\bibinfo {year}
  {2012})}\BibitemShut {NoStop}%
\bibitem [{\citenamefont {Coulais}\ \emph {et~al.}(2015)\citenamefont
  {Coulais}, \citenamefont {Overvelde}, \citenamefont {Lubbers}, \citenamefont
  {Bertoldi},\ and\ \citenamefont {van Hecke}}]{Coulais_Metabeam}%
  \BibitemOpen
  \bibfield  {author} {\bibinfo {author} {\bibfnamefont {C.}~\bibnamefont
  {Coulais}}, \bibinfo {author} {\bibfnamefont {J.~T.~B.}\ \bibnamefont
  {Overvelde}}, \bibinfo {author} {\bibfnamefont {L.~A.}\ \bibnamefont
  {Lubbers}}, \bibinfo {author} {\bibfnamefont {K.}~\bibnamefont {Bertoldi}}, \
  and\ \bibinfo {author} {\bibfnamefont {M.}~\bibnamefont {van Hecke}},\
  }\bibfield  {title} {\enquote {\bibinfo {title} {Discontinuous buckling of
  wide beams and metabeams},}\ }\href
  {http://link.aps.org/doi/10.1103/PhysRevLett.115.044301} {\bibfield
  {journal} {\bibinfo  {journal} {Phys. Rev. Lett.}\ }\textbf {\bibinfo
  {volume} {115}},\ \bibinfo {pages} {044301} (\bibinfo {year}
  {2015})}\BibitemShut {NoStop}%
\bibitem [{\citenamefont {Coulais}\ \emph {et~al.}(2017)\citenamefont
  {Coulais}, \citenamefont {Sounas},\ and\ \citenamefont
  {Alù}}]{Coulais_nonreci}%
  \BibitemOpen
  \bibfield  {author} {\bibinfo {author} {\bibfnamefont {C.}~\bibnamefont
  {Coulais}}, \bibinfo {author} {\bibfnamefont {D.}~\bibnamefont {Sounas}}, \
  and\ \bibinfo {author} {\bibfnamefont {Andrea}\ \bibnamefont {Alù}},\
  }\bibfield  {title} {\enquote {\bibinfo {title} {Static non-reciprocity in
  mechanical metamaterials},}\ }\href {\doibase 10.1038/nature21044} {\bibfield
   {journal} {\bibinfo  {journal} {Nature}\ }\textbf {\bibinfo {volume}
  {542}},\ \bibinfo {pages} {461--464} (\bibinfo {year} {2017})}\BibitemShut
  {NoStop}%
\bibitem [{\citenamefont {Kane}\ and\ \citenamefont
  {Lubensky}(2014)}]{KaneLubensky}%
  \BibitemOpen
  \bibfield  {author} {\bibinfo {author} {\bibfnamefont {C.~L.}\ \bibnamefont
  {Kane}}\ and\ \bibinfo {author} {\bibfnamefont {T.~C.}\ \bibnamefont
  {Lubensky}},\ }\bibfield  {title} {\enquote {\bibinfo {title} {Topological
  boundary modes in isostatic lattices},}\ }\href
  {http://dx.doi.org/10.1038/nphys2835} {\bibfield  {journal} {\bibinfo
  {journal} {Nat. Phys.}\ }\textbf {\bibinfo {volume} {10}},\ \bibinfo {pages}
  {39--45} (\bibinfo {year} {2014})}\BibitemShut {NoStop}%
\bibitem [{\citenamefont {Chen}\ \emph {et~al.}(2014)\citenamefont {Chen},
  \citenamefont {Upadhyaya},\ and\ \citenamefont {Vitelli}}]{Vitelli_solitons}%
  \BibitemOpen
  \bibfield  {author} {\bibinfo {author} {\bibfnamefont {B.~G.}\ \bibnamefont
  {Chen}}, \bibinfo {author} {\bibfnamefont {N.}~\bibnamefont {Upadhyaya}}, \
  and\ \bibinfo {author} {\bibfnamefont {V.}~\bibnamefont {Vitelli}},\
  }\bibfield  {title} {\enquote {\bibinfo {title} {Nonlinear conduction via
  solitons in a topological mechanical insulator},}\ }\href {\doibase
  10.1073/pnas.1405969111} {\bibfield  {journal} {\bibinfo  {journal} {Proc.
  Natl. Ac. Sc. U. S. A.}\ }\textbf {\bibinfo {volume} {111}},\ \bibinfo
  {pages} {13004--9} (\bibinfo {year} {2014})}\BibitemShut {NoStop}%
\bibitem [{\citenamefont {Huber}(2016)}]{Huber_reviewtopological}%
  \BibitemOpen
  \bibfield  {author} {\bibinfo {author} {\bibfnamefont {S.~D.}\ \bibnamefont
  {Huber}},\ }\bibfield  {title} {\enquote {\bibinfo {title} {Topological
  mechanics},}\ }\href {\doibase 10.1038/nphys3801} {\bibfield  {journal}
  {\bibinfo  {journal} {Nat. Phys.}\ }\textbf {\bibinfo {volume} {12}},\
  \bibinfo {pages} {621--623} (\bibinfo {year} {2016})}\BibitemShut {NoStop}%
\bibitem [{\citenamefont {Meeussen}\ \emph {et~al.}(2016)\citenamefont
  {Meeussen}, \citenamefont {Paulose},\ and\ \citenamefont
  {Vitelli}}]{Meeussen_gears}%
  \BibitemOpen
  \bibfield  {author} {\bibinfo {author} {\bibfnamefont {A.~S.}\ \bibnamefont
  {Meeussen}}, \bibinfo {author} {\bibfnamefont {J.}~\bibnamefont {Paulose}}, \
  and\ \bibinfo {author} {\bibfnamefont {V.}~\bibnamefont {Vitelli}},\
  }\bibfield  {title} {\enquote {\bibinfo {title} {Geared topological
  metamaterials with tunable mechanical stability},}\ }\href
  {http://link.aps.org/doi/10.1103/PhysRevX.6.041029} {\bibfield  {journal}
  {\bibinfo  {journal} {Phys. Rev. X}\ }\textbf {\bibinfo {volume} {6}},\
  \bibinfo {pages} {041029} (\bibinfo {year} {2016})}\BibitemShut {NoStop}%
\bibitem [{\citenamefont {Waitukaitis}\ \emph {et~al.}(2015)\citenamefont
  {Waitukaitis}, \citenamefont {Menaut}, \citenamefont {Chen},\ and\
  \citenamefont {van Hecke}}]{Waitukaitis_origami}%
  \BibitemOpen
  \bibfield  {author} {\bibinfo {author} {\bibfnamefont {S.}~\bibnamefont
  {Waitukaitis}}, \bibinfo {author} {\bibfnamefont {R.}~\bibnamefont {Menaut}},
  \bibinfo {author} {\bibfnamefont {B.~G.}\ \bibnamefont {Chen}}, \ and\
  \bibinfo {author} {\bibfnamefont {M.}~\bibnamefont {van Hecke}},\ }\bibfield
  {title} {\enquote {\bibinfo {title} {Origami multistability: from single
  vertices to metasheets},}\ }\href {\doibase 10.1103/PhysRevLett.114.055503}
  {\bibfield  {journal} {\bibinfo  {journal} {Phys. Rev. Lett.}\ }\textbf
  {\bibinfo {volume} {114}},\ \bibinfo {pages} {055503} (\bibinfo {year}
  {2015})}\BibitemShut {NoStop}%
\bibitem [{\citenamefont {Silverberg}\ \emph {et~al.}(2015)\citenamefont
  {Silverberg}, \citenamefont {Na}, \citenamefont {Evans}, \citenamefont {Liu},
  \citenamefont {Hull}, \citenamefont {Santangelo}, \citenamefont {Lang},
  \citenamefont {Hayward},\ and\ \citenamefont
  {Cohen}}]{Silverberg_squaretwist}%
  \BibitemOpen
  \bibfield  {author} {\bibinfo {author} {\bibfnamefont {J.~L.}\ \bibnamefont
  {Silverberg}}, \bibinfo {author} {\bibfnamefont {J.~H.}\ \bibnamefont {Na}},
  \bibinfo {author} {\bibfnamefont {A.~A.}\ \bibnamefont {Evans}}, \bibinfo
  {author} {\bibfnamefont {B.}~\bibnamefont {Liu}}, \bibinfo {author}
  {\bibfnamefont {T.~C.}\ \bibnamefont {Hull}}, \bibinfo {author}
  {\bibfnamefont {C.~D.}\ \bibnamefont {Santangelo}}, \bibinfo {author}
  {\bibfnamefont {R.~J.}\ \bibnamefont {Lang}}, \bibinfo {author}
  {\bibfnamefont {R.~C.}\ \bibnamefont {Hayward}}, \ and\ \bibinfo {author}
  {\bibfnamefont {I.}~\bibnamefont {Cohen}},\ }\bibfield  {title} {\enquote
  {\bibinfo {title} {Origami structures with a critical transition to
  bistability arising from hidden degrees of freedom},}\ }\href {\doibase
  10.1038/nmat4232} {\bibfield  {journal} {\bibinfo  {journal} {Nat. Mater.}\
  }\textbf {\bibinfo {volume} {14}},\ \bibinfo {pages} {389--93} (\bibinfo
  {year} {2015})}\BibitemShut {NoStop}%
\bibitem [{\citenamefont {Coulais}\ \emph {et~al.}(2016)\citenamefont
  {Coulais}, \citenamefont {Teomy}, \citenamefont {de~Reus}, \citenamefont
  {Shokef},\ and\ \citenamefont {van Hecke}}]{Coulais_metacube}%
  \BibitemOpen
  \bibfield  {author} {\bibinfo {author} {\bibfnamefont {C.}~\bibnamefont
  {Coulais}}, \bibinfo {author} {\bibfnamefont {E.}~\bibnamefont {Teomy}},
  \bibinfo {author} {\bibfnamefont {K.}~\bibnamefont {de~Reus}}, \bibinfo
  {author} {\bibfnamefont {Y.}~\bibnamefont {Shokef}}, \ and\ \bibinfo {author}
  {\bibfnamefont {M.}~\bibnamefont {van Hecke}},\ }\bibfield  {title} {\enquote
  {\bibinfo {title} {Combinatorial design of textured mechanical
  metamaterials},}\ }\href {\doibase 10.1038/nature18960} {\bibfield  {journal}
  {\bibinfo  {journal} {Nature}\ }\textbf {\bibinfo {volume} {535}},\ \bibinfo
  {pages} {529--531} (\bibinfo {year} {2016})}\BibitemShut {NoStop}%
\bibitem [{\citenamefont {Dudte}\ \emph {et~al.}(2016)\citenamefont {Dudte},
  \citenamefont {Vouga}, \citenamefont {Tachi},\ and\ \citenamefont
  {Mahadevan}}]{Maha_curvature}%
  \BibitemOpen
  \bibfield  {author} {\bibinfo {author} {\bibfnamefont {L.~H.}\ \bibnamefont
  {Dudte}}, \bibinfo {author} {\bibfnamefont {E.}~\bibnamefont {Vouga}},
  \bibinfo {author} {\bibfnamefont {T.}~\bibnamefont {Tachi}}, \ and\ \bibinfo
  {author} {\bibfnamefont {L.}~\bibnamefont {Mahadevan}},\ }\bibfield  {title}
  {\enquote {\bibinfo {title} {Programming curvature using origami
  tessellations},}\ }\href {\doibase 10.1038/nmat4540} {\bibfield  {journal}
  {\bibinfo  {journal} {Nat. Mater.}\ }\textbf {\bibinfo {volume} {15}},\
  \bibinfo {pages} {583--8} (\bibinfo {year} {2016})}\BibitemShut {NoStop}%
\bibitem [{\citenamefont {Overvelde}\ \emph {et~al.}(2016)\citenamefont
  {Overvelde}, \citenamefont {de~Jong}, \citenamefont {Shevchenko},
  \citenamefont {Becerra}, \citenamefont {Whitesides}, \citenamefont {Weaver},
  \citenamefont {Hoberman},\ and\ \citenamefont
  {Bertoldi}}]{Overvelde_origami1}%
  \BibitemOpen
  \bibfield  {author} {\bibinfo {author} {\bibfnamefont {J.~T.}\ \bibnamefont
  {Overvelde}}, \bibinfo {author} {\bibfnamefont {T.~A.}\ \bibnamefont
  {de~Jong}}, \bibinfo {author} {\bibfnamefont {Y.}~\bibnamefont {Shevchenko}},
  \bibinfo {author} {\bibfnamefont {S.~A.}\ \bibnamefont {Becerra}}, \bibinfo
  {author} {\bibfnamefont {G.~M.}\ \bibnamefont {Whitesides}}, \bibinfo
  {author} {\bibfnamefont {J.~C.}\ \bibnamefont {Weaver}}, \bibinfo {author}
  {\bibfnamefont {C.}~\bibnamefont {Hoberman}}, \ and\ \bibinfo {author}
  {\bibfnamefont {K.}~\bibnamefont {Bertoldi}},\ }\bibfield  {title} {\enquote
  {\bibinfo {title} {A three-dimensional actuated origami-inspired
  transformable metamaterial with multiple degrees of freedom},}\ }\href
  {\doibase 10.1038/ncomms10929} {\bibfield  {journal} {\bibinfo  {journal}
  {Nat. Commun.}\ }\textbf {\bibinfo {volume} {7}},\ \bibinfo {pages} {10929}
  (\bibinfo {year} {2016})}\BibitemShut {NoStop}%
\bibitem [{\citenamefont {Overvelde}\ \emph {et~al.}(2017)\citenamefont
  {Overvelde}, \citenamefont {Weaver}, \citenamefont {Hoberman},\ and\
  \citenamefont {Bertoldi}}]{Overvelde_origami2}%
  \BibitemOpen
  \bibfield  {author} {\bibinfo {author} {\bibfnamefont {J.~T.}\ \bibnamefont
  {Overvelde}}, \bibinfo {author} {\bibfnamefont {J.~C.}\ \bibnamefont
  {Weaver}}, \bibinfo {author} {\bibfnamefont {C.}~\bibnamefont {Hoberman}}, \
  and\ \bibinfo {author} {\bibfnamefont {K.}~\bibnamefont {Bertoldi}},\
  }\bibfield  {title} {\enquote {\bibinfo {title} {Rational design of
  reconfigurable prismatic architected materials},}\ }\href {\doibase
  10.1038/nature20824} {\bibfield  {journal} {\bibinfo  {journal} {Nature}\
  }\textbf {\bibinfo {volume} {541}},\ \bibinfo {pages} {347--352} (\bibinfo
  {year} {2017})}\BibitemShut {NoStop}%
\bibitem [{\citenamefont {Silverberg}\ \emph {et~al.}(2014)\citenamefont
  {Silverberg}, \citenamefont {Evans}, \citenamefont {McLeod}, \citenamefont
  {Hayward}, \citenamefont {Hull}, \citenamefont {Santangelo},\ and\
  \citenamefont {Cohen}}]{Silverberg_miura}%
  \BibitemOpen
  \bibfield  {author} {\bibinfo {author} {\bibfnamefont {J.~L.}\ \bibnamefont
  {Silverberg}}, \bibinfo {author} {\bibfnamefont {A.~A.}\ \bibnamefont
  {Evans}}, \bibinfo {author} {\bibfnamefont {L.}~\bibnamefont {McLeod}},
  \bibinfo {author} {\bibfnamefont {R.~C.}\ \bibnamefont {Hayward}}, \bibinfo
  {author} {\bibfnamefont {T.}~\bibnamefont {Hull}}, \bibinfo {author}
  {\bibfnamefont {C.~D.}\ \bibnamefont {Santangelo}}, \ and\ \bibinfo {author}
  {\bibfnamefont {I.}~\bibnamefont {Cohen}},\ }\bibfield  {title} {\enquote
  {\bibinfo {title} {Using origami design principles to fold reprogrammable
  mechanical metamaterials},}\ }\href
  {http://www.sciencemag.org/content/345/6197/647.abstract
  http://science.sciencemag.org/content/sci/345/6197/647.full.pdf} {\bibfield
  {journal} {\bibinfo  {journal} {Science}\ }\textbf {\bibinfo {volume}
  {345}},\ \bibinfo {pages} {647--650} (\bibinfo {year} {2014})}\BibitemShut
  {NoStop}%
\bibitem [{\citenamefont {Florijn}\ \emph {et~al.}(2014)\citenamefont
  {Florijn}, \citenamefont {Coulais},\ and\ \citenamefont {van
  Hecke}}]{Florijn_biholar}%
  \BibitemOpen
  \bibfield  {author} {\bibinfo {author} {\bibfnamefont {B.}~\bibnamefont
  {Florijn}}, \bibinfo {author} {\bibfnamefont {C.}~\bibnamefont {Coulais}}, \
  and\ \bibinfo {author} {\bibfnamefont {M.}~\bibnamefont {van Hecke}},\
  }\bibfield  {title} {\enquote {\bibinfo {title} {Programmable mechanical
  metamaterials},}\ }\href
  {http://link.aps.org/doi/10.1103/PhysRevLett.113.175503} {\bibfield
  {journal} {\bibinfo  {journal} {Phys. Rev. Lett.}\ }\textbf {\bibinfo
  {volume} {113}},\ \bibinfo {pages} {175503} (\bibinfo {year}
  {2014})}\BibitemShut {NoStop}%
\bibitem [{\citenamefont {Bertoldi}\ \emph {et~al.}(2010)\citenamefont
  {Bertoldi}, \citenamefont {Reis}, \citenamefont {Willshaw},\ and\
  \citenamefont {Mullin}}]{Bertoldi_poisson}%
  \BibitemOpen
  \bibfield  {author} {\bibinfo {author} {\bibfnamefont {K.}~\bibnamefont
  {Bertoldi}}, \bibinfo {author} {\bibfnamefont {P.~M.}\ \bibnamefont {Reis}},
  \bibinfo {author} {\bibfnamefont {S.}~\bibnamefont {Willshaw}}, \ and\
  \bibinfo {author} {\bibfnamefont {T.}~\bibnamefont {Mullin}},\ }\bibfield
  {title} {\enquote {\bibinfo {title} {Negative poisson's ratio behavior
  induced by an elastic instability},}\ }\href
  {http://dx.doi.org/10.1002/adma.200901956} {\bibfield  {journal} {\bibinfo
  {journal} {Adv. Mater.}\ }\textbf {\bibinfo {volume} {22}},\ \bibinfo {pages}
  {361--366} (\bibinfo {year} {2010})}\BibitemShut {NoStop}%
\bibitem [{\citenamefont {Milton}\ and\ \citenamefont
  {Cherkaev}(1995)}]{Milton_tensors}%
  \BibitemOpen
  \bibfield  {author} {\bibinfo {author} {\bibfnamefont {G.~W.}\ \bibnamefont
  {Milton}}\ and\ \bibinfo {author} {\bibfnamefont {A.~V.}\ \bibnamefont
  {Cherkaev}},\ }\bibfield  {title} {\enquote {\bibinfo {title} {Which
  elasticity tensors are realizable?}}\ }\href
  {http://dx.doi.org/10.1115/1.2804743} {\bibfield  {journal} {\bibinfo
  {journal} {J. Eng. Mater. Technol.}\ }\textbf {\bibinfo {volume} {117}},\
  \bibinfo {pages} {483--493} (\bibinfo {year} {1995})}\BibitemShut {NoStop}%
\bibitem [{\citenamefont {Lechenault}\ \emph {et~al.}(2014)\citenamefont
  {Lechenault}, \citenamefont {Thiria},\ and\ \citenamefont
  {Adda-Bedia}}]{Lechenault_OriLengthscale}%
  \BibitemOpen
  \bibfield  {author} {\bibinfo {author} {\bibfnamefont {F.}~\bibnamefont
  {Lechenault}}, \bibinfo {author} {\bibfnamefont {B.}~\bibnamefont {Thiria}},
  \ and\ \bibinfo {author} {\bibfnamefont {M.}~\bibnamefont {Adda-Bedia}},\
  }\bibfield  {title} {\enquote {\bibinfo {title} {Mechanical response of a
  creased sheet},}\ }\href {\doibase 10.1103/PhysRevLett.112.244301} {\bibfield
   {journal} {\bibinfo  {journal} {Phys. Rev. Lett.}\ }\textbf {\bibinfo
  {volume} {112}},\ \bibinfo {pages} {244301} (\bibinfo {year}
  {2014})}\BibitemShut {NoStop}%
\bibitem [{\citenamefont {Landau}\ and\ \citenamefont
  {Lifshitz}()}]{Landau_Lifschitz}%
  \BibitemOpen
  \bibfield  {author} {\bibinfo {author} {\bibfnamefont {L.~D.}\ \bibnamefont
  {Landau}}\ and\ \bibinfo {author} {\bibfnamefont {E.~M.}\ \bibnamefont
  {Lifshitz}},\ }\href@noop {} {\emph {\bibinfo {title} {Theory of
  elasticity}}}\ (\bibinfo  {publisher} {Pergamon},\ \bibinfo {address}
  {Oxford, UK})\BibitemShut {NoStop}%
\bibitem [{\citenamefont {Rocks}\ \emph {et~al.}(2017)\citenamefont {Rocks},
  \citenamefont {Pashine}, \citenamefont {Bischofberger}, \citenamefont
  {Goodrich}, \citenamefont {Liu},\ and\ \citenamefont
  {Nagel}}]{Nagel_allostery}%
  \BibitemOpen
  \bibfield  {author} {\bibinfo {author} {\bibfnamefont {J.~W.}\ \bibnamefont
  {Rocks}}, \bibinfo {author} {\bibfnamefont {N.}~\bibnamefont {Pashine}},
  \bibinfo {author} {\bibfnamefont {I.}~\bibnamefont {Bischofberger}}, \bibinfo
  {author} {\bibfnamefont {C.~P.}\ \bibnamefont {Goodrich}}, \bibinfo {author}
  {\bibfnamefont {A.~J.}\ \bibnamefont {Liu}}, \ and\ \bibinfo {author}
  {\bibfnamefont {S.~R.}\ \bibnamefont {Nagel}},\ }\bibfield  {title} {\enquote
  {\bibinfo {title} {Designing allostery-inspired response in mechanical
  networks},}\ }\href {\doibase 10.1073/pnas.1612139114} {\bibfield  {journal}
  {\bibinfo  {journal} {Proc. Natl. Ac. Sc. U. S. A.}\ }\textbf {\bibinfo
  {volume} {114}},\ \bibinfo {pages} {2520--2525} (\bibinfo {year}
  {2017})}\BibitemShut {NoStop}%
\bibitem [{\citenamefont {Yan}\ \emph {et~al.}(2017)\citenamefont {Yan},
  \citenamefont {Ravasio}, \citenamefont {Brito},\ and\ \citenamefont
  {Wyart}}]{Wyart_allostery}%
  \BibitemOpen
  \bibfield  {author} {\bibinfo {author} {\bibfnamefont {L.}~\bibnamefont
  {Yan}}, \bibinfo {author} {\bibfnamefont {R.}~\bibnamefont {Ravasio}},
  \bibinfo {author} {\bibfnamefont {C.}~\bibnamefont {Brito}}, \ and\ \bibinfo
  {author} {\bibfnamefont {M.}~\bibnamefont {Wyart}},\ }\bibfield  {title}
  {\enquote {\bibinfo {title} {Architecture and coevolution of allosteric
  materials},}\ }\href {\doibase 10.1073/pnas.1615536114} {\bibfield  {journal}
  {\bibinfo  {journal} {Proc. Natl. Ac. Sc. U. S. A.}\ }\textbf {\bibinfo
  {volume} {114}},\ \bibinfo {pages} {2526--2531} (\bibinfo {year}
  {2017})}\BibitemShut {NoStop}%
\end{thebibliography}

\begin{thebibliography}{21}
\expandafter\ifx\csname natexlab\endcsname\relax\def\natexlab#1{#1}\fi
\expandafter\ifx\csname bibnamefont\endcsname\relax
  \def\bibnamefont#1{#1}\fi
\expandafter\ifx\csname bibfnamefont\endcsname\relax
  \def\bibfnamefont#1{#1}\fi
\expandafter\ifx\csname citenamefont\endcsname\relax
  \def\citenamefont#1{#1}\fi
\expandafter\ifx\csname url\endcsname\relax
  \def\url#1{\texttt{#1}}\fi
\expandafter\ifx\csname urlprefix\endcsname\relax\def\urlprefix{URL }\fi
\providecommand{\bibinfo}[2]{#2}
\providecommand{\eprint}[2][]{\url{#2}}


  \bibitem[{\citenamefont {Grima} and \citenamefont{Evans}(2000)}]{Grima_squares}
\bibinfo{author}{\bibfnamefont{J.~N.}~\bibnamefont{Grima}} \bibnamefont{and}
  \bibinfo{author}{\bibfnamefont{K.~E.}~\bibnamefont{Evans}},
  \emph{\bibinfo{title}{Auxetic behavior  from rotating squares}},
  \bibinfo{journal}{J. Mater. Sc. Lett.} \textbf{\bibinfo{volume}{19}},
  \bibinfo{pages}{1563--1565} (\bibinfo{year}{2000}).
  
  \bibitem[{\citenamefont{Coulais at al. }(2017)\citenamefont{Coulais at al. }}]{Coulais_nonreci}
\bibinfo{author}{\bibfnamefont{C.}~\bibnamefont{Coulais}},  \bibinfo{author}{\bibfnamefont{D.}~\bibnamefont{Sounas}} \bibnamefont{and}
  \bibinfo{author}{\bibfnamefont{A.}~\bibnamefont{Al\`u}},
  \emph{\bibinfo{title}{Static non-reciprocity in mechanical metamaterials}},
  \bibinfo{journal}{Nature} \textbf{\bibinfo{volume}{542}},
  \bibinfo{pages}{461-464} (\bibinfo{year}{2017}).

 
  
\end{thebibliography}
\end{document}